\documentstyle[12pt]{article}
\newcommand{\fslash}[1]{\mbox{$\!\not\!#1$}}
\newcommand{\Bar}[1]{\overline{#1}}
\newcommand{\threeqBS}{3qBS}

\newlength{\Textwidth}
\newlength{\Tatescale}
\newcommand{\Tate}{\rule{0cm}{\Tatescale}}
\newcommand{\Tatetate}{\rule{0cm}{1.5\Tatescale}}
\newcommand{\Rule}[1]{\rule{0em}{#1}}

\newcommand{\Zero}{^{[0]}}
\newcommand{\tozero}{\mbox{\bf(!)}}
\newcommand{\tildebar}[1]{\tilde{\Bar{#1}}}

\newcommand{\Tr}{\mbox{Tr}}
\newcommand{\Hs}{\hspace*{1em}}
\newcommand{\Bs}{\hspace*{-0.5em}}
\newcommand{\zr}[1]{\mbox{\hspace*{#1em}}}
\newcommand{\ID}{\mbox{{\sf 1}\zr{-0.14}\rule{0.04em}{1.55ex}\zr{0.1}}}

\newcommand{\Fig}[1]{{\rm Fig}.\ref{#1}}
\newcommand{\Eq}[1]{Eq.({\ref{#1}})}
\newcommand{\Appendix}[1]{Appendix {\ref{#1}}}

\newcommand{\Section}[1]{Section {\ref{#1}}}

\title{{\begin{flushright}
\large FAU--TP3--00/6 \hfill nucl-th/0004063
\end{flushright}}
\mbox{ }\\[1cm]
Goldberger--Treiman Relation and $g_{\pi{NN}}$\\
from the Three Quark BS/Faddeev Approach\\ in the NJL Model}
\author{ Noriyoshi ISHII\\
	Institute for Theoretical Physics III,\\
	University of Erlangel-N\"urnberg,\\
	Staudtstrasse 7, D-91058, Erlangen, GERMANY\\
	E-mail: ishii@theorie3.physik.uni-erlangen.de\\
	Phone: +49 +9131 85 28466\\
	FAX: +49 +9131 85 27704}
\date{ }
\begin{document}
\maketitle
\begin{abstract}
A  systematic evaluation method of  matrix  elements of quantum one-body
operators     between    nucleon      states   in     the    three-quark
Bethe-Salpeter(\threeqBS{})/Faddeev    approach   in  the   NJL model is
reviewed.  We do not confine ourselves to a particular truncation scheme
of the \threeqBS{} kernel.  One of our main aims  is to derive a general
condition  to be imposed  on a  given BS kernel   in order that the PCAC
relation   is satisfied  correctly.   We  apply  this condition  to some
particular  \threeqBS{} kernels.  We  numerically calculate $g_{\pi NN}$
in  the \threeqBS{} /Faddeev approach  to estimate the  violation of the
Goldberger-Treiman/PCAC relation  due to the  UV-regularization  scheme.
We finally consider the non-vanishing current  quark mass effects on the
Goldberger-Treiman relation.
\end{abstract}

\smallskip
\leftline{\bf PACS Classification Codes:} 12.39.-x, 12.39, Fe, 12.39.Ki
\smallskip

\smallskip
\leftline{\bf        Keywords:}  Nucleon,     Chiral  Ward     identity,
Goldberger-Treiman     Relation,  $g_{\pi  NN}$,    Relativistic Faddeev
Equation, Nambu-Jona-Lasinio (NJL) Model.
\smallskip
\newpage
\section{Introduction}

The Nambu-Jona-Lasinio(NJL)  model  is    a quantum  field   theoretical
effective quark model based on QCD.  While the confinement, which is one
of  the  most important   properties  of the  low energy   QCD,   is not
incorporated, it  is the simplest quantum  field theoretical quark model
respecting the chiral symmetry  and providing us with  explicit examples
how the  spontaneous breaking of  the chiral symmetry plays an important
role    in    hadronic   phenomena\cite{nambu},    in   particular,  the
interpretation of  pion, K, and  $\eta$ as Nambu-Goldstone modes.  Based
on  these  advantages,  not  only  mesons  but  also  baryons have  been
extensively studied  in the NJL model\cite{vogl.weise,njl}.  The studies
of baryons are mainly classified into the following two categories: (C1)
the  relativistic  mean    field approach\cite{soliton}, and   (C2)  the
relativistic      three-quark   Bethe-Salpeter(\threeqBS{}     )/Faddeev
approach\cite{reindhardt,ishii1,ishii2,ishii3,asami}.  Since   the   NJL
model is a second quantized field  theory, it is, in principle, possible
to consider the effects of $q\Bar{q}$ excitations in  the nucleon.  This
is one of the  most interesting  targets  in the studies of  the nucleon
structure beyond the non-relativistic constituent quark models.  In this
respect, on the one hand, the meanfield approach incorporates $q\Bar{q}$
effects through the mesonic hedgehog   fields leading to the   solitonic
picture of the nucleon,  where, however, the manifest Lorentz covariance
is unfortunately missing.  On  the other hand, the  \threeqBS{} /Faddeev
approach, which respects the manifest  Lorentz covariance leading to the
quark-diquark   picture  of   baryons\cite{diquarks}, approximates   the
\threeqBS{}  kernel according  to the  ladder  truncation scheme,  which
prevents   us from studying    non-trivial $q\Bar{q}$ effects beyond the
excitations of  the ``three-quark RPA vacuum''\cite{ishii4}(the backward
diagrams included in the ladder approximation).

It is thus necessary to extend the \threeqBS{} kernel  in order to study
the  non-trivial $q\Bar{q}$ effects  in a Lorentz  covariant manner, and
therefore attempts  have been  (and  are) made  to go beyond  the ladder
truncation  scheme \cite{lorenz1}.     To  extend  the  NJL  \threeqBS{}
kernels, chiral  symmetry imposes important constraints.  In particular,
it would be very useful to have  a criterion which  tells us which kinds
of kernels  lead to the PCAC relation\footnote{\label{footnote.1}We mean
by ``PCAC relation''
in the sense of   \Eq{desired.relation}.}  correctly.  One of  the  main
aims of this paper is to derive such a criterion and to provide explicit
examples of its application to some of the existing \threeqBS{} kernels.
We  will first review how  matrix elements of quantum one-body operators
are evaluated systematically in the \threeqBS{} framework by introducing
classical external fields as a technical tool.  Note that the evaluation
of the bound  state matrix elements in the  \threeqBS{} framework is not
obvious beyond the   diagramatic  argument.  We  will  provide a  direct
formula  to evaluate  the  matrix   element in   terms of the    Faddeev
framework.  We then consider which properties  of the \threeqBS{} kernel
are required  to satisfy the PCAC  relation correctly.  We  consider the
relevant Feynman diagrams to  calculate the matrix elements for  several
\threeqBS{} kernels,  and apply  our   criterion  to these   \threeqBS{}
kernels.  Although we  are going to restrict our  attention only  to the
PCAC  case,  these considerations  themselves can  be  extended to other
cases straightforwardly such as the electromagnetic current, the isospin
current, the baryon number  current, etc ---actually,  the PCAC  case is
the most complicated  one\footnote{The proof of the electromagnetic Ward
identity  in  the  Faddeev approach  to  the  NJL  model in  the  ladder
truncation scheme can be found in ref.\cite{asami}, which was applied to
the quark-diquark  model in \cite{lorenz2}.   Although these authors did
not use  the external field method, the  results are consistent with our
formalism.  A systematic approach using the external field method can be
found   in ref.\cite{sasha}.}.

Even if the truncation scheme is consistent with  the PCAC relation, the
UV regularization  schemes, which have been  adopted so far in practical
numerical calculations in the NJL model,  usually, spoil it.  It is thus
worth while to evaluate $g_{\pi NN}$ and $g_{A}$  in the chiral limit in
order to estimate  explicitly the  violation of the   Goldberger-Treiman
relation due to the UV regularization scheme.  We will see that with the
Euclidean sharp  cut-off the violation  is up  to 4 \%.   The  extracted
value of $g_{\pi NN}$ is 13.2.  We will also estimate the PCAC violation
due to the cutoff scheme off the chiral limit, which is again up to 4\%.
We  will evaluate the off-shell $g_{\pi  NN} (\equiv \tilde g_{\pi NN})$
with the  single-pole    dominance approximation for  the   form  factor
$h_A(q^2)$.  We obtain   a very reasonable value  $\tilde   g_{\pi NN} =
13.5$.  However $g_A$ is 6 \% larger than the experimental one.  We will
also consider the effect of the  non-vanishing current quark mass on the
Goldberger-Treiman   relation, and find   that   it is in  the  opposite
direction  than expected from the  experimental values.  We will analyze
this by including also the  effect of the $q\bar  q$ interaction in  the
color  singlet  iso-vector  axial-vector channel,  providing an analytic
expression for the  deviation from  the Goldberger-Treiman relation   by
assuming  that  the vacuum is  approximated  according to the mean-field
approximation.   We   will find  that  the   effect  of this  additional
$q\bar{q}$ channel works into the unwanted direction.  We will see that,
to  resolve  this problem,  it is necessary   either  to improve the gap
equation  for  the vacuum beyond  the   mean field  approximation or  to
evaluate the on-shell $g_{\pi NN}$.

\section{Bound State Matrix Elements and the Criterion (Sufficient Condition)
for the PCAC Relation}
\label{section.2}

We  consider effective quark   Lagrangians  with global $SU(2)_f  \times
SU(3)_c$ symmetries:
\begin{equation}
	{\cal L}
=
	\Bar{\psi}\left( i \fslash{\partial} - m_0 \right)\psi
+	{\cal L}_{\rm I},
\label{lagrangian}
\end{equation}
where  $m_0$ is the current quark  mass, and $\psi$ and $\Bar{\psi}$ are
the quark  bispinor fields.   ${\cal L}_{\rm I}$  is a   local, chirally
symmetric four-fermionic interaction    Lagrangian of  NJL  type,  i.e.,
$\displaystyle     {\cal  L}_{\rm   I}      = \sum_\Gamma     g_{\Gamma}
\left(\Bar{\psi}\Gamma\psi\right)^2$, where  $\Gamma$  is a  matrix with
Dirac, isospin, and  color indices.  Examples are  the original NJL type
${\cal           L}_I           =        g\left((\Bar\psi\psi)^2       -
(\Bar\psi\gamma_5\vec\tau\psi)^2\right)$  and      the    color  current
interaction            type                ${\cal         L}_I         =
g\left(\Bar\psi\gamma^{\mu}{\lambda^a\over2}\psi\right)^2$, etc.    (For
detail,     see    \Appendix{convensions.of.the.njl.model}.1.)     Since
interaction Lagrangians of this  type can all be dealt  with in the same
manner, we will not confine ourselves to a particular one.

To evaluate a matrix element in the  \threeqBS{} framework is not really
straight forward.  This  is  because what  is directly obtained  in  the
\threeqBS{} framework  is not the  nucleon eigen-ket $|N\rangle$ but the
\threeqBS{}  amplitude\footnote{The \threeqBS{}   amplitude    is  often
referred   to as the   ``wave  function''  due  to  historical  reasons.
However, we  prefer  to call it as   ``\threeqBS{} amplitude'' to  avoid
unnecessary  confusions\cite{itzykson}.},   which   itself is a   matrix
element of the  type $\langle 0|T\psi\psi\psi|N\rangle$.  We might  thus
suspect that a huge amount of information, which is contained in the ket
vector $|N\rangle$,  could be missing in the  \threeqBS{} amplitude.  In
the  relativistic  Faddeev  framework,  the situation  is slightly  more
complicated ---there are   no immediate representations of the   Faddeev
amplitude  in terms of  the  canonical  operator  formalism\footnote{The
Faddeev  amplitude  provides the   same amount  of informations  as  the
\threeqBS{}         amplitude     does,    as        we      show     in
\Appendix{appendix.faddeev.equation}.}.  Therefore, we will first review
how to evaluate  a matrix element in the  \threeqBS{} framework based on
the canonical operator formalism.

To this end, it is convenient to introduce space-time dependent external
fields:
\begin{equation}
    v_{\mu}(x)      =  v^a_{\mu}(x){\tau^a\over  2},
\Hs
    a_{\mu}(x)  = a^a_{\mu}(x){\tau^a\over 2},
\Hs
    m(x) = s(x)  + ip^a(x)\gamma_5{\tau^a\over 2},
\end{equation}
where $v^a_{\mu}(x)$ is an  isovector vector field, $a^a_{\mu}(x)$ is an
isovector axial-vector field, $s(x)$   is a scalar isoscalar  field, and
$p^a(x)$ is a  pseudo-scalar isovector  field.  (The iso-spin index  $a$
runs over 1, 2, 3.)  The following  Lagrangian density characterizes how
these   external  fields couple   to the  associated  quantized one-body
operators:
\begin{eqnarray}
	{\cal L}^{[e]}
&=&
	\Bar{\psi}
	\left(
		i \fslash{\partial} - \fslash{v} - \fslash{a}\gamma_5 - m
	\right)
	\psi + {\cal L}_{\rm I}
\\
&=&
	\Bar\psi i\fslash{\partial} \psi + {\cal L}_I
	- v^a_{\mu}V^{a\mu} - a^a_{\mu}A^{a\mu} - \Bar{\psi}m\psi,
\nonumber
\end{eqnarray}
where $\displaystyle V^{a\mu}  \equiv \Bar{\psi}\gamma^{\mu}{\tau^a\over
2}\psi$          and      $\displaystyle       A^{a\mu}           \equiv
\Bar{\psi}\gamma^{\mu}\gamma_5{\tau^a\over   2}\psi$.  The  superscripts
$[e]$ or  $[0]$ will be used  to indicate quantities  in the presence or
absence,  respectively,  of the  external  fields  $v,a,m -  m_0$.   For
simplicity, we include the current quark mass $m_0$ in the definition of
the  external field $m$.  We assume  that these external  fields, $v, a,
m$, are localized  in  space-time,  i.e.,  they vanish except  for  some
finite space-time region.  After performing the necessary manipulations,
we  will eventually set $a(x)\equiv v(x)\equiv   0$ and $m\equiv m_0$ in
the whole space-time.  This limit will be refereed to as the ``vanishing
external fields limit''.

The \threeqBS{} equation    in the presence of  the   external fields is
derived from the Schwinger-Dyson(SD) equation\cite{itzykson} by means of
the same technique as for vanishing external fields:
\begin{equation}
	G^{[e]}
=
	G^{[e]}_0
+
	G^{[e]}_0
	V^{[e]}
	G^{[e]},
\label{3q.bs.equation.in.the.presence.of.external.field}
\end{equation}
where $V^{[e]}$ is the sum of  2PI and 3PI interactions, and $G^{[e]}_0$
is an anti-symmetric combination of products  of three constituent quark
propagators in the presence of the external  fields.  For simplicity, we
denote
\begin{equation}
	K^{[e]} \equiv  G^{[e]}_0    V^{[e]}.
\label{k=g0v}
\end{equation}
To avoid cumbersome notations, we  often adopt the operator notation  to
suppress the explicit integration symbol $\displaystyle \int d^4 x_1 d^4
x_2 d^4 x_3$,  the explicit space-times coordinates $(x_1,x_2,x_3)$  and
the explicit indices of the quark fields.

Since the external fields are localized  in space--time, a nucleon state
``propagates''   freely in the  ``past''.  Therefore,   $G^{[e]}$ has an
asymptotic  initial nucleon   pole.   The residue  at  this nucleon pole
satisfies the following homogeneous \threeqBS{} equation in the presence
of the external fields:
\begin{equation}
	\Psi[N_{\alpha}^a(\vec{P})]^{[e]}
=
	K^{[e]}\Psi[N_{\alpha}^a(\vec{P})]^{[e]},
\label{homogeneous.equation}
\end{equation}
where $\Psi[N_{\alpha}^a(\vec{P})]^{[e]}$ is the \threeqBS{}   amplitude
for the nucleon in the presence of the external  fields with the isospin
$a$,        helicity $\alpha$  and  asymptotic    four    momentum $P  =
(E_N(\vec{P}^2),\vec{P})$     with   $E_N(\vec{P}^2)   = \sqrt{m_N^2   +
\vec{P}^2}$ ($m_N$ : nucleon mass).  In terms  of the canonical operator
formalism, it is expressed as:
\begin{eqnarray}
\lefteqn{\Psi[N_{\alpha}^a(\vec{P})]^{[e]}(x_1,x_2,x_3)}
\label{canonical}
\\
&=&
	\Bs
	\begin{array}[t]{l}	\displaystyle
		\left\langle 0 \left| \Tate\right.\right.
		T\psi(x_1)\psi(x_2)\psi(x_3)
	\\	\displaystyle
	\times
		\exp -i\int d^4x
		\left[
			v^a_{\mu}(x)V^{a\mu}(x)
			+ a^a_{\mu}(x)A^{a\mu}(x)
			+ \Bar{\psi}\left( m(x) - m_0 \right)\psi
		\right]
	\left.\left.\Tate\right| N_{\alpha}^a(\vec{P}) \right\rangle.
	\end{array}
\nonumber
\end{eqnarray}
$\left|\left.\Tate    0\right\rangle\right.$    and   $\left|\left.\Tate
N_{\alpha}^a(\vec P)\right\rangle\right.$ are the vacuum and the nucleon
state    vectors,   respectively.    They   are   eigenstates  of    the
``unperturbed''   Hamiltonian $H^{[0]}$   obtained by   setting  $v=a=0,
m=m_0$.  We adopt the covariant normalizations:
\begin{equation}
	\left\langle 0  \left|\Tate\right. 0\right\rangle
=
	1,
\Hs
	\left\langle N_{\alpha'}^{a'}(\vec{P}')  \left| \Tate
	\right. N_{\alpha}^{a}(\vec{P}) \right\rangle
=
	{E_N(\vec P^2)\over m_N}
	(2\pi)^3 \delta^{(3)}(\vec{P}' - \vec{P})
	\delta_{\alpha'\alpha}\delta_{a'a}.
\label{covariant.normalization}
\end{equation}
Note that  quantized  operators are   represented in  the  ``interaction
picture''      in      \Eq{canonical},    where   the      ``unperturbed
Hamiltonian''\footnote{ Note that  $H^{{}[0]}$  is the full   Heisenberg
Hamiltonian in the absence  of the external  fields.}  is  $H^{[0]}$ and
the ``perturbing Hamiltonian'' is
\begin{equation}
	H_I
\equiv
	\int d^3 x
	\left(
		v^a_{\mu}(x)V_{a\mu}(x)
	+	a^a_{\mu}(x)A^{a\mu}(x)
	+	\Bar{\psi}(x)\left( m(x) - m_0 \right)\psi(x)
	\right).
\end{equation}
We next  consider  the matrix element  $\langle N|A^{b}_{\mu}|N\rangle$.
Applying  the functional derivative $\displaystyle{i\delta\over   \delta
a^a_{\mu}(x)}$ to both sides of \Eq{homogeneous.equation}, we obtain the
following relation in the vanishing external field limit,
\begin{equation}
	\left(
	\displaystyle{
		i \delta\Psi[N_{\alpha}^a(\vec{P})]^{[e]}
	\over 	\delta a^{b}_{\mu}(x)
	}
	\right)\Zero
=
	\left(\displaystyle{
		i \delta K^{[e]}\over \delta a^b_{\mu}(x)
	}\right)\Zero
	\Psi[N_{\alpha}^a(\vec{P})]^{[0]}
	+
	K^{[0]}
	\left(\displaystyle{
		i \delta \Psi[N_{\alpha}^a(\vec{P})]^{[e]}
		\over \delta a^b_{\mu}(x)
	}\right)\Zero.
\label{intermediate}
\end{equation}
The expressions    for   the  ``\threeqBS{}  amplitudes''   entering  in
\Eq{intermediate} are given in terms of the canonical operator formalism
as follows:
\begin{eqnarray}
	\Psi[N_{\alpha}^{a}(\vec{P})]^{[0]}(x_1,x_2,x_3)
&=&
	\left\langle 0 \left| \Tate
		T\psi(x_1)\psi(x_2)\psi(x_3)
	\right| N_{\alpha}^a(\vec{P}) \right\rangle
\label{3q.bs.amplitude}
\\
\nonumber
	\Bar\Psi[N_{\alpha}^{a}(\vec{P})]^{[0]}(x_1,x_2,x_3)
&=&
	\left\langle N^a_\alpha(\vec{P}) \left| \Tate
		T\Bar\psi(x_1)\Bar\psi(x_2)\Bar\psi(x_3)
	\right| 0 \right\rangle
\\
\nonumber
\left(
	{i \delta \Psi[N_{\alpha}^a(\vec{P})]^{[e]}(x_1,x_2,x_3)
	\over \delta a^b_{\mu}(x)}
\right)\Zero
&=&
	\left\langle 0 \left| \Tate
		T\psi(x_1)\psi(x_2)\psi(x_3) A^{b\mu}(x)
	\right| N_{\alpha}^a(\vec{P})\right\rangle.
\end{eqnarray}
By using the canonical operator  analysis,  i.e., by inserting into  the
Green's function  the complete  set  relation in   the baryon number   1
sector:
\begin{equation}
    \ID
=
    \sum_{a\alpha} \int {d^3 p \over (2\pi)^3}
    {m_N \over E_N(\vec p^2)}
    \left|\left. \Tate N^a_{\alpha}(\vec p) \right\rangle\right.
    \left\langle\left. \Tate N^a_{\alpha}(\vec p) \right|\right.
+
    \cdots,
\end{equation}
and by using the Fourier  transform of  the step  function  $\theta(t)$,
we see that\footnote{The argument is similar to the one given in p.92 in
\cite{itzykson}.}  the product $\Psi^{[0]}\Bar\Psi^{[0]}$ provides   the
residue of the Green's function at the nucleon pole, i.e.,
\begin{equation}
    G^{[0]}
=
    \sum_{a\alpha}
    \int {d^4p\over (2\pi)^4}
    {m_N \over E_N(\vec p^2)}
    {i \over p_0 - E_N(\vec p^2) + i\epsilon}
    \Psi[N^a_{\alpha}(\vec p)]^{[0]}
    \Bar\Psi[N^a_\alpha(\vec p)]^{[0]}
+
    \cdots
\end{equation}
In \Appendix{appendix.faddeev.equation},   we explain how  to obtain and
normalize these  two  \threeqBS{} amplitudes based   on the relativistic
Faddeev  framework.
In \Appendix{appendix.faddeev.equation},    we prove that   the residues
$\Psi^{[0]}$ and   $\Bar\Psi^{[0]}$  satisfy  the following  \threeqBS{}
equation    in       the    absence   of      the    external  fields(cf
\Eq{homogeneous.equation})
\begin{equation}
\label{homogeneous.\threeqBS{} .equation.without.external.field}
    K^{[0]}
    \Psi^{[0]}[N^a_{\alpha}(\vec P)]
=
    \Psi^{[0]}[N^a_{\alpha}(\vec P)],
\Hs
    \tildebar{\Psi}^{[0]}[N^a_{\alpha}(\vec P)]
    K^{[0]}
=
    \tildebar{\Psi}^{[0]}[N^a_{\alpha}(\vec P)],
\end{equation}
where     a  tilde  is   used  to    indicate an   ``amputation'', i.e.,
$\displaystyle   \tildebar{\Psi}[N^a_\alpha(\vec       p)]^{[0]}       =
\Bar\Psi[N^a_\alpha(\vec    p)]^{[0]}     G_0^{[0]-1}$.   By rearranging
\Eq{intermediate}, we obtain the following relation:
\begin{equation}
\left({
	i \delta \Psi[N_{\alpha}^a(\vec{P})]^{[e]} \over \delta a^b_{\mu}(x)
}\right)\Zero
=
	{1\over 1 - K^{[0]}}
	\left({
		i \delta K^{[e]} \over \delta a^b_{\mu}(x)
	}\right)\Zero
	\Psi[N_{\alpha}^a(\vec{P})]^{[0]}.
\label{main.equation}
\end{equation}
On the one hand, near the nucleon pole the resolvent is given as
\begin{eqnarray}
    {1\over 1 - K^{[0]}}
&=&
    G^{[0]}
    G^{[0]-1}_{0}
\\\nonumber
&\simeq&
	\sum_{a,\alpha}
	\int {d^4 p\over (2\pi)^4}
	{m_N \over E_N(\vec{p}^2)}
	{i \over p^0 - E_N(\vec{p}^2) + i\epsilon}
	\Psi[N^a_\alpha(\vec{p})]^{[0]}
	\tildebar{\Psi}[N^a_\alpha(\vec{p})]^{[0]}.
\end{eqnarray}
The l.h.s.  of \Eq{main.equation}, on the  other hand, is given near the
nucleon pole as follows:
\begin{eqnarray}
\lefteqn{
	\left(
		{i \delta \Psi[N^a_\alpha(\vec{P})]^{[e]}
	\over
		\delta a^b_{\mu}(x)}
	\right)^{[0]}
}
\\
\nonumber
&\simeq&
	\sum_{a'\alpha'}
	\int {d^4p\over(2\pi)^4}
	{m_N\over E_N(\vec{p}^2)}
	{i\over p^0 - E_N(\vec{p}^2) + i\epsilon}
	\left\langle 0 \left| \Tate
		T\psi\psi\psi
	\right| N^{a'}_{\alpha'}(\vec{p}) \right\rangle
	\left\langle N^{a'}_{\alpha'}(\vec{p}) \left|
		A^b_{\mu}(x)
	\right| N^a_\alpha(\vec{P}) \right\rangle.
\\
\nonumber
&=&
	\sum_{a'\alpha'}
	\int {d^4p\over(2\pi)^4}
	{m_N\over E_N(\vec{p}^2)}
	{i\over p'^0 - E_N(\vec{p}^2) + i\epsilon}
	\Psi[N^{a'}_{\alpha'}(\vec{p})]^{[0]}
	\left\langle N^{a'}_{\alpha'}(\vec{p}) \left|
		A^b_{\mu}(x)
	\right| N^a_\alpha(\vec{P}) \right\rangle.
\end{eqnarray}
By comparing both sides of \Eq{main.equation}, we are left with:
\begin{equation}
	\left\langle N^{a'}_{\alpha'}(\vec{P}') \left|
		A^b_{\mu}(x)
	\right| N^a_{\alpha}(\vec{P}) \right\rangle
=
	\tildebar{\Psi}[N^{a'}_{\alpha'}(\vec{P}')]^{[0]}
	\left(
		{i \delta K^{[e]}\over \delta a^b_{\mu}(x)}
	\right)\Zero
	\Psi[N^a_{\alpha}(\vec{P})]^{[0]}.
\label{matrix.element.1}
\end{equation}
A similar consideration leads to the following relation:
\begin{equation}
	\left\langle N^{a'}_{\alpha'}(\vec{P}') \left| \Tate
		i\Bar{\psi}(x)\gamma_5{\tau^b\over 2}\psi(x)
	\right| N^{a}_{\alpha}(\vec{P}) \right\rangle
=
	\tildebar{\Psi}[N^{b'}_{\alpha'}(\vec{P}')]^{[0]}
	\left(\displaystyle{
		i \delta K^{[e]}\over \delta p^b(x) 
	}\right)\Zero
	\Psi[N^a_{\alpha}(\vec{P})]^{[0]}.
\label{matrix.element.2}
\end{equation}
A combination of these two relations provides us with:
\begin{eqnarray}
\lefteqn{
	\left\langle N^{a'}(\vec{P}') \left| \Tate
		\partial_{\mu}A^{b\mu}(x)
		- 2im_0 \Bar{\psi}(x)\gamma_5\tau^b\psi(x)
	\right|N^{a}(\vec{P})\right\rangle
}
\label{matrix.element.3}
\\
&=&
	\tildebar{\Psi}[N^{a'}_{\alpha'}(\vec{P}')]^{[0]}
	\left(
		\left(
		\partial_{\mu}
		\displaystyle{i \delta \over \delta a_{\mu}^b(x)}
		- 2m_0
		\displaystyle{i \delta \over \delta p^b(x)}
		\right)
		K^{[e]}
	\right)\Zero
	\Psi[N^a_\alpha(\vec{P})]^{[0]}.
\nonumber
\end{eqnarray}
These  expressions show that the   \threeqBS{} amplitudes provide enough
informations to  evaluate  these matrix elements.  Here   the ``relative
time''  dependence of the  \threeqBS{}   amplitude surely plays a   very
important role.    At any rate, we   should keep in mind  that, although
bound   state  matrix  elements are  obtained    by sandwiching one-body
operators in the canonical  operator   formalism, it is these   two-body
operators  shown on the  r.h.s.  of \Eq{matrix.element.3} that should be
sandwiched in the \threeqBS{} framework\footnote{If 3PI interactions are
included in the \threeqBS{} kernel, there appear three-body operators in
addition to the two-body operators in the \threeqBS{} framework.}, which
leads to crucial  differences\cite{sasha}.  The corresponding formula to
evaluate  the bound state  matrix elements  in the  Faddeev framework is
derived in \Appendix{appendix.faddeev.equation}.    Note that it  is not
hard  to  extend   these   arguments   beyond   the  NJL  model.

If the full \threeqBS{} kernel were at our disposal, the chiral symmetry
of the original Lagrangian should directly lead to  the PCAC relation in
the \threeqBS{} framework.   However, because the  \threeqBS{} kernel is
in practice truncated  such as to be  manageable, it may happen that the
truncation scheme spoils the chiral symmetry.  We are thus interested in
the criterion to decide which \threeqBS{}  kernel gives rise to the PCAC
relation correctly.  For this purpose, it is convenient to introduce the
local ``axial'' gauge transformation of the external fields as follows:
\begin{eqnarray}
    i\fslash{\partial}
    - \fslash{v}^{(\omega)}(x)
    - \fslash{a}^{(\omega)}(x)\gamma_5
    - m^{(\omega)}
&=&
    \Omega(x)
    \left(	\Rule{\Tatescale}
	i\fslash{\partial}
	- \fslash{v}(x)
	- \fslash{a}(x)\gamma_5
	- m
    \right)
    \Omega(x),
\label{def.gauge.transform}
\\\nonumber
    \Omega(x)
&=&
    e^{-i\gamma_5\omega(x)};
\Hs
    \omega(x)= \omega^a(x){\tau^a\over 2},
\end{eqnarray}
where $v^{(\omega)}$,  $a^{(\omega)}$ and  $m^{(\omega)}$ are  the gauge
images of $v$, $a$ and $m$, respectively.  We assume that $\omega(x)$ is
also   localized    in    space-time.   Infinitesimally,    these  gauge
transformations are expressed as follows:
\begin{eqnarray}
    {\delta v^a_{\mu}(y)\over \delta\omega^c(x)}
&=&
    \epsilon_{abc}a^b_{\mu}(y)\delta^{(4)}(x - y)
\\
    {\delta a^a_{\mu}(y)\over \delta\omega^c(x)}
&=&
    \delta_{ac}
    {\partial\over \partial x^{\mu}}
    \delta^{(4)}(x - y)
+
    \epsilon_{abc}v^b_{\mu}(y)\delta^{(4)}(x - y)
\nonumber
\\
    {\delta s(y)\over \delta \omega^c(x)}
&=&
    {1\over 2}p^c(y)\delta^{(4)}(x - y)
\nonumber
\\
    {\delta p^b(y)\over\delta\omega^c(x)}
&=&
    - 2\delta_{bc}s(y)\delta^{(4)}(x - y),
\nonumber
\end{eqnarray}
where $\epsilon_{abc}$   is   the totally  anti-symmetric  tensor.   Let
$F[v,a,m]$ be   a   functional.  We define   the   functional derivative
$\displaystyle{\delta\over   \delta\omega^c(x)}$  through the  following
relation:
\begin{equation}
    { \delta F \over \delta \omega^c(x) }
\equiv
    {
	\delta F[v^{(\omega)},a^{(\omega)},m^{(\omega)}]
    \over
	\delta \omega^c(x)
    }.
\end{equation}
By using   the    chain  rule, $\displaystyle{i     \delta\over   \delta
\omega^c(x)}$ is  expressed  as   a linear  combination   of  functional
derivatives with respect to the external fields:
\begin{eqnarray}
\lefteqn{
	\displaystyle{i\delta\over \delta \omega^c(x)}
}
\\
&=&
	\int d^4y
	\left[
		{\delta a^b_{\mu}(y)\over \delta\omega^c(x)}
		{i \delta \over \delta a^b_{\mu}(y)}
		+
		{\delta v^b_{\mu}(y)\over \delta\omega^c(x)}
		{i \delta \over \delta v^b_{\mu}(y)}
		+
		{\delta s(y)\over\delta\omega^c(x)}
		{i \delta\over \delta s(y)}
		+
		{\delta p^b(y)\over\delta\omega^c(x)}
		{i \delta\over \delta p^b(y)}
	\right]
\nonumber
\\
&=&
	\partial_{\mu}{i \delta\over \delta a^c_{\mu}(x)}
	-
	\epsilon_{abc}v^a_{\mu}(x)
	{i \delta\over \delta a^b_{\mu}(x)}
	-
	\epsilon_{abc}a^a_{\mu}(x)
	{i \delta\over \delta v^b_{\mu}(x)}
	+
	{1\over 2} p^c(x)
	{i \delta \over \delta s(x)}
	-
	2 s(x)
	{i \delta \over \delta p^c(x)}.
\nonumber
\end{eqnarray}
In particular, we have the  following relation in the vanishing external
field limit:
\begin{equation}
	\left(
		{i\delta K^{[e^{(\omega)}]}
	\over
		\delta\omega^b(x)}
	\right)
	\Zero
=
	\partial_{\mu}
	\left(
		{i\delta K^{[e]}\over \delta a^b_{\mu}(x)}
	\right)
	\Zero
	- 2 m_0
	\left(
		{i\delta K^{[e]}\over \delta p^b(x)}
	\right)
	\Zero,
\label{eq.23}
\end{equation}
where  $[e^{(\omega)}]$    denotes     the gauge   transformed    fields
$[v^{(\omega)},a^{(\omega)},m^{(\omega)}]$.  Now we state the sufficient
condition for the validity of the PCAC relation as follows:

\smallskip
\leftline{\bf Sufficient  Condition:}
\smallskip
If a  \threeqBS{}   kernel behaves  in  the following manner under  a gauge
transformation $\Omega(x)$:
\begin{eqnarray}
\lefteqn{
	K^{[e^{(\omega)}]}(x_1,x_2,x_3;y_1,y_2,y_3)
}
\label{eq.24}
\\
&=&
    \left(\Tate
	\Omega(x_1)^{-1} \otimes \Omega(x_2)^{-1} \otimes \Omega(x_3)^{-1}
    \right)
    K^{[e]}
    \left(\Tate
	\Omega(y_1) \otimes \Omega(y_2) \otimes \Omega(y_3)
    \right),
\nonumber
\end{eqnarray}
then the  PCAC   relation  is satisfied.

Our ``criterion'' is obtained from \Eq{eq.24} by applying the functional
derivative with respect to the  gauge transformation on both sides,  and
using \Eq{eq.23}.  We are left with  the following chiral Ward identity,
which is the ``criterion'' for the validity of the PCAC relation:\\
\smallskip
{\bf Criterion:} (The Chiral Ward Identity for the \threeqBS{}  Kernel)
\begin{eqnarray}
\lefteqn{
	\left[
	\left(
		\partial_{\mu}
		\displaystyle{i\delta\over \delta a^{b}_{\mu}(x)}
		- 2 m_0
		\displaystyle{i\delta\over \delta p^b(x)}
	\right)
	K^{[e]}
	\right]\Zero
	(x_1,x_2,x_3;y_1,y_2,y_3)
}
\label{ward.id}
\\
&=&
    \Bs
    \begin{array}[t]{l}	\displaystyle
	\rule{0ex}{2ex}
        K^{[0]}
        \rule{0ex}{3ex}
        \left[ \Tate
	    (\tau^b\gamma_5\!\otimes\!1\!\otimes\!1)\delta^{(4)}(x-y_1)\;
	    +\; (1\!\otimes\!\tau^b\gamma_5\!\otimes\!1)\delta^{(4)}(x-y_2)\;
	    +\; (1\!\otimes\!1\!\otimes\!\tau^b\gamma_5)\delta^{(4)}(x-y_3)
	\right]
    \\	\displaystyle \Rule{1.5\Tatescale}
	-
        \left[
	    (\tau^b\gamma_5\!\otimes\!1\!\otimes\!1)\delta^{(4)}(x-x_1) \;
	    +\; (1\!\otimes\!\tau^b\gamma_5\!\otimes\!1)\delta^{(4)}(x-x_2)\;
	    +\; (1\!\otimes\!1\!\otimes\!\tau^b\gamma_5)\delta^{(4)}(x-x_3)
	\right]
	K^{[0]}.
    \end{array}
\nonumber
\end{eqnarray}

To see   that  this indeed   leads to the   PCAC   relation, we sandwich
\Eq{ward.id}    between  $\Psi[N^{a'}_{\alpha'}(\vec{P}')]^{[0]}$    and
$\tildebar{\Psi}[N^a_\alpha(\vec{P})]^{[0]}$            and          use
\Eq{homogeneous.\threeqBS{}    .equation.without.external.field}     and
\Eq{matrix.element.3} to get
\begin{equation}
	\left\langle N^{a'}_{\alpha'}(\vec{P}')    \left|  \Tate
		\partial_\mu  A^{b\mu}(x)
	-	2 m_0  i \Bar{\psi}(x)\gamma_5\tau^b\psi(x)
	\right| N^a_\alpha(\vec{P}) \right\rangle
=
	0,
\label{desired.relation}
\end{equation}
which is the desired relation.

We  should comment   here on  the   reason  why we  called \Eq{eq.24}  a
``sufficient  condition''.  If  we   try    to extend our   method    to
$SU(3)_L\times SU(3)_R$-QCD
in   the  presence  of external    chiral   gauge fields, the   relation
(\ref{eq.24}) is no longer valid due to the existence of the non-Abelian
anomaly\footnote{This anomaly should not be confused with the $U_{A}(1)$
anomaly, i.e.,   the Abelian anomaly.}  \cite{pokorski}.  Note, however,
that, even if there is such a non-Abelian anomaly, because the anomalous
contributions are  polynomials of at  least second order in the external
chiral gauge fields, the infinitesimal form (the criterion \Eq{ward.id})
is still valid.  However, if more than two functional derivatives of the
external  chiral gauge fields are  involved, one must pay full attention
to  the  non-Abelian anomaly even for   the  infinitesimal form.  At any
rate, the chiral $SU(2)_L\times SU(2)_R$ case, which is of most interest
to us here, is known to be anomaly free.

\section{Feynman Diagrams}

The aim of this section  is to see  which kinds of Feynman diagrams  are
involved   in    the  expression   \Eq{matrix.element.1}   of   $\langle
N|A_{\mu}^b(x)|N\rangle$,    before   applying  the criterion/sufficient
condition to particular \threeqBS{} kernels.

We  first consider the constituent  quark propagator $S_F^{[e]}(x,y)$ in
the presence of the external fields.   The self energy is, in principle,
obtained  from the  sum of  1PI diagrams.   However, in practice,  it is
approximated in the mean field (Hartree-Fock) treatment, as expressed by
the following gap equation:
\begin{eqnarray}
    iS_F^{[e]}(x,y)
&=&
    iS_{0;F}^{[e]}(x,y)
\label{gap.equation}
\\\nonumber
&-&
    \sum_\Gamma 2 i g_{\Gamma}^{(q\Bar q)}
    \int d^4 z
    iS_{0;F}^{[e]}(x,z)
    \Gamma iS_F^{[e]}(z,y)
    \Tr\left(\Gamma i S_F^{[e]}(z,z) \right).
\end{eqnarray}
$g_{\Gamma}^{(q\Bar q)}$ are the   effective coupling constants  in  the
$q\bar      q$     channels,         which       are   defined        in
\Appendix{convensions.of.the.njl.model}.1.  $S_{0;F}^{[e]}(x,y)$ is  the
current quark   propagator, which   is  defined through    the following
relation:
\begin{equation}
	\left(
		i\fslash{\partial}^{(x)}
	-	\fslash{a}(x)\gamma_5
	-	\fslash{v}(x)
	-	m(x)
	\right)
	S_{0;F}^{[e]}(x,y)
=
	\delta^{(4)}(x-y).
\end{equation}
The  solution to  \Eq{gap.equation}  is  obtained  as a  self-consistent
solution to the following equations:
\begin{eqnarray}
	S_F^{[e]}(x,y)
&=&
	\left(
		i\fslash{\partial} - \fslash{a}\gamma_5
	-	\fslash{v}
	-	m - \Sigma^{[e]}
	\right)^{-1}(x,y)
\label{self.consistent}
\\
	\Sigma^{[e]}(z)
&\equiv&
	\sum_\Gamma 2ig_\Gamma^{(q\Bar q)}
	\Tr\left(
		S_F^{[e]}(z,z) \Gamma
	\right) \Gamma.
\nonumber
\end{eqnarray}
$\Sigma^{[e]}(z)$ is the self-energy  of  the constituent  quark.   Now,
several comments are in order.  (1) Due  to the presence of the external
fields, there  is no translational symmetry any  more, and therefore the
self-energy   depends on the space-time point   $z$.  (The dependence on
just  a  single  space-time  coordinate  $z$ is  due  to the  mean field
approximation.)  (2) Non-vanishing external fields may  lead to not only
a    non-vanishing scalar    condensate,    but  also   a  non-vanishing
pseudoscalar, vector,  axial   vector  condensates.   However,   in  the
vanishing external field  limit, we have  $m_0 + \Sigma^{[0]}  = M$ (the
constituent  quark    mass).     (3)   Because  the   NJL     model   is
non-renormalizable, it is necessary to introduce  a UV regularization in
all loop integrals.  Hereafter, whenever a loop integral is encountered,
the  integration is  understood  as  a regularized   one.   (4) If  this
regularization respects    the  chiral symmetry,   $S_F^{[e]}(x,y)$  and
$\Sigma^{[e]}(z)$  should  behave in   the following way:
\begin{eqnarray}
	S_F^{[e^{(\omega)}]}(x,y)
&=&
	\Omega(x)^{-1}\;
	S_F^{[e]}(x,y)\;
	\Omega(y)^{-1}
\label{gauge.transformation.propagator}
\\
	\Sigma^{[e^{(\omega)}]}(z)
&=&
	\Omega(z)\;
	\Sigma^{[e]}(z)\;
	\Omega(z).
\nonumber
\end{eqnarray}
For later convenience,  the   transformation of  $S_F$ is  depicted   in
\Fig{fig1}.

All diagrammatically truncated    \threeqBS{} kernels $K^{[e]}$   can be
expressed as the    product  of several constituent  quark   propagators
$S_F^{[e]}$ and several elementary local vertices $\Gamma$.  Because, in
our case, these elementary local vertices do not  depend on the external
fields\footnote{\protect\label{derivative.coupling}If         derivative
couplings are involved in  the  interaction Lagrangian, our argument  is
straightforwardly extended by  replacing ordinary derivatives  by chiral
covariant derivatives.  In  this case, $\delta/\delta a_{\mu}^b(x)$ also
hits  the argument  $\Gamma$   in $K[\Gamma^{[e]};S_F^{[e]}]$.  This  is
essential to extend  the arguments given in  the next section.  In  this
case, it is not simply $A^{b\mu}(x)$ but the conserved current operator,
that is inserted in the r.h.s.  in  the 3rd line in \Eq{3q.bs.amplitude}
---  $A^{b\mu}(x)$  fails  to be a   conserved  current in this  case.},
$K^{[e]}$ depends on $v,a$   and $m$ only   through $S_F^{[e]}$.   In  a
symbolic notation,  this   may  be denoted   as $K^{[e]}   =   K[\Gamma;
S_F^{[e]}]$,  and  the  functional   derivatives  are then  symbolically
expressed by using the chain rule as follows:
\begin{equation}
	\left[
	{i \delta K^{[e]}\over \delta a_{\mu}^b(x)}
	\right]\Zero
=
	\sum_{\alpha\alpha'}
	\int d^4 z d^4 z'
	\left[
	{
		i \delta {S_F}^{[e]}_{\alpha\alpha'}(z,z')
		\over
		\delta a_{\mu}^b(x)}
	\right]\Zero
	\left(
	{
		\delta K[ \Gamma;S_F^{[0]}]
		\over
		\delta {S_F^{[0]}}_{\alpha\alpha'}(z,z')
	}
	\right),
\label{eq.31}
\end{equation}
where  $\alpha,\alpha'$ are triples  of  the Dirac, iso-spin,  and color
indices.  This   relation implies that  the general  rule  to obtain the
functional  derivative of the   \threeqBS{}  kernel is to   replace each
constituent quark propagator in turn by  its derivative, and then sum up
the resulting terms.    Now  all  we  need  is  $\displaystyle   {\delta
S_F^{[e]}/  \delta    a^b_{\mu}(x)}$, which   is   obtained  by applying
$\displaystyle {\delta/ \delta a^b_{\mu}(x)}$ to \Eq{self.consistent} as
follows:
\begin{eqnarray}
    \left[
	{\delta S_F^{[e]}(x,y) \over \delta a_{\mu}^b(z)}
    \right]\Zero
&=&
    \Bs
    \begin{array}[t]{l}	\displaystyle
	S_F\Zero(x,z)
	\left(\Tate \gamma^{\mu}\gamma_5 {\tau^b \over 2} \right)
	S_F\Zero(z,y)
    \\	\displaystyle
    +	
	\int d^4 z'
	S_F\Zero(x,z')
	\left[
	    {\delta \Sigma^{[e]}(z')\over \delta a_{\mu}^b(z)}
        \right]\Zero
	S_F\Zero(z',y).
    \end{array}
\label{functional.derivative.propagator}
\\
    \left[
    {
    	\delta \Sigma^{[e]}(z')
    \over
	\delta a_{\mu}^b(z)
    }
    \right]
    \Zero
&=&
    \sum_\Gamma 2ig_{\Gamma}^{(q\Bar q)}
    \Tr\left(
	\left[
	{
	    \delta S^{[e]}_F(z',z')
	    \over
		\delta a_{\mu}^b(z)
	}
	\right]
	\Zero
	\Gamma
    \right)
    \Gamma.
\end{eqnarray}
By inserting the first relation into the second, we obtain the following
closed equation for $\displaystyle {\delta \Sigma/ \delta a_{\mu}^b}$:
\begin{equation}
    \left[
    {
	\delta \Sigma^{[e]}(z')
    \over
	\delta a_{\mu}^b(z)
    }
    \right]^{[0]}
=
    \Bs
    \begin{array}[t]{l}	\displaystyle
	\sum_{\Gamma}
	2ig_{\Gamma}^{(q\Bar q)}
	\mbox{Tr}\left(
	    S^{[0]}(z',z)
	    \left(\Tate \gamma^{\mu}\gamma_5 {\tau^b\over 2}\right)
	    S^{[0]}(z,z')
	    \Gamma
	\right)
	\Gamma
    \\\displaystyle\Rule{1.5\Tatescale}
    +
	\int d^4 z''
	\sum_{\Gamma} 2ig_{\Gamma}^{(q\Bar q)}
	\mbox{Tr}\left(
	    S^{[0]}(z',z'')
	    \left[
	        {\delta \Sigma^{[e]}(z'') \over \delta a_{\mu}^b(z)}
	    \right]^{[0]}
	    S^{[0]}(z'',z')
	    \Gamma
	\right)
	\Gamma.
    \end{array}
\label{closed.eq.for.dsigma.da}
\end{equation}
For simplicity, we  adopt\footnote{The non-vanishing effective $q\bar q$
coupling constant in the  iso-vector  axial-vector channel $g_{\rm  ax}$
does not  change  the   chiral symmetry  properties\cite{nambu}.}    the
effective $q\Bar q$  coupling   constant in the iso-vector   axialvector
channel $g_{\rm ax} = 0$.  (For the precise meaning of $g_{\rm ax}$, see
\Appendix{convensions.of.the.njl.model}.  We  will discuss   the general
case,    i.e.,         $g_{\rm         ax}      \neq        0$,       in
\Appendix{appendix.non-vanishing.gax.gtv.index}.)   We  can parameterize
our solution as follows:
\begin{equation}
    \int d^4 z'
    e^{iq(z' - z)}
    \left[{\delta \Sigma^{[e]}(z') \over \delta a^{b}_{\mu}(z)}\right]^{[0]}
=
    \tilde H(q^2) \left(q_{\mu}\gamma_5 {\tau^b\over 2} \right),
\end{equation}
and   from \Eq{closed.eq.for.dsigma.da}  we  see   that  $\tilde H(q^2)$
satisfies the following equation:
\begin{equation}
    \tilde H(q^2)
=
    -2ig_{\pi} \Pi_{5A}(q^2) - 2ig_{\pi} \Pi_{55}(q^2) \tilde H(q^2),
\end{equation}
where $\Pi_{5A}(q^2)$ and $\Pi_{55}(q^2)$ are the bubble integrals which
are  defined in \Appendix{convensions.of.the.njl.model}.2.  The solution
is expressed as a geometric series:
\begin{equation}
    \tilde H(q^2)
=
    {-2ig_{\pi} \Pi_{5A}(q^2) \over 1 + 2ig_{\pi}\Pi_{55}(q^2)}.
\label{eq.37}
\end{equation}
$\displaystyle {  \delta  S_F / \delta   a_{\mu}^b(z)}$  is depicted  in
\Fig{fig2}.(a).  The two diagrams on  the r.h.s.  correspond to the  two
terms in  \Eq{functional.derivative.propagator}, respectively, where the
second one is proportional to  $q^{\mu}\tilde H(q^2)$ and has the pionic
pole.   Note  that, whenever the    quark propagator  is  obtained as  a
non-trivial  solution to a  self-consistent equation which  leads to the
chiral symmetry breaking, $\delta   S_F/\delta a_{\mu}^b$ always   has a
non-trivial ``mesonic part'' $\tilde H(q^2)$.

Now we can consider particular  \threeqBS{}  kernels and the  associated
diagrams relevant for the matrix element calculation.  Our first example
is  the  ladder  truncated  \threeqBS{}  kernel, which   is depicted  in
\Fig{fig3}(a).  Solid  lines represent the constituent quark propagator.
A slash  indicates amputation of the  constituent quark propagator.  The
interaction  strengths in   the  various $qq$ channels   are obtained by
applying  the Fierz identity  to    the interaction Lagrangian.     (See
\Appendix{convensions.of.the.njl.model}.)  In  this     case, \Eq{eq.31}
leads to the diagram in \Fig{fig3}(b), where  the insertion on the quark
line        indicated by  the   ``$\otimes$''     has   been defined  in
\Fig{fig2}. (Diagrams with the same topologies are omitted.)

The  second example  is the $qq$   interaction involving the exchange of
$q\bar q$ pairs in the t-channel (``meson exchange interaction''), which
is  depicted in  \Fig{fig4}(a).  In  this  case, \Eq{eq.31} leads to the
diagrams   depicted in \Fig{fig4}(b).  We  see  that, in addition to the
coupling  of  the external field   to  the external  quark  propagators,
couplings  to internal quark  propagators  are also involved (the second
diagram), which are often referred  to as the ``meson exchange current''
contributions.

It is straightforward to extend these considerations to more complicated
and realistic  cases.  A moral  is that, once  the \threeqBS{} kernel is
specified, a unique set   of Feynman diagrams  exists to   determine the
matrix element of the axial  vector current.

\section{Applications of the Criterion/Sufficient Condition}

The aims   of this section  are to  see  whether the \threeqBS{} kernels
presented  in   the      previous   section  satisfy   the     criterion
\Eq{ward.id}/sufficient condition \Eq{eq.24},   and to provide  examples
how to use the general considerations of Section 2 practically.

We first consider the chiral transformation properties of the elementary
local vertices in  the Lagrangian.   The  global chiral  symmetry of the
interaction Lagrangian $\displaystyle {\cal  L}_I = \sum_\Gamma g_\Gamma
(\Bar\psi\Gamma\psi)^2$ implies the following identity:
\begin{equation}
	\sum_\Gamma g_\Gamma
	\left(
		e^{i\gamma_5 \tau^b \Theta^b}
		\Gamma
		e^{i\gamma_5 \tau^b \Theta^b}
	\right)_{ij}
	\left(
		e^{i\gamma_5 \tau^b \Theta^b}
		\Gamma
		e^{i\gamma_5 \tau^b \Theta^b}
	\right)_{kl}
=
	\sum_\Gamma g_\Gamma \Gamma_{ij}\Gamma_{kl}.
\end{equation}
Since  ${\cal L}_I$ is   a contact  interaction  without any  derivative
terms, this   identity  remains valid      even if $\Theta^b$    acquires
space--time dependence, i.e.,
\begin{equation}
    \sum_\Gamma g_\Gamma
    \left(\Tate
	\Omega(x)
	\Gamma
	\Omega(x)
    \right)_{ij}
    \left(\Tate
	\Omega(x)
	\Gamma
	\Omega(x)
    \right)_{kl}
=
    \sum_\Gamma g_\Gamma \Gamma_{ij}\Gamma_{kl}.
\label{chiral.invariance}
\end{equation}
Actually,  these  $q\Bar{q}$  interactions  consist of  several ``closed
chiral multiplet  sectors'',      and   the  identity   of   the    type
\Eq{chiral.invariance} holds  in  each   such  sector  separately.   For
example, the  $q\bar  q$ interaction in  the  $\pi$ and $\sigma$ mesonic
channel  forms a closed  chiral  multiplet under the  local axial  gauge
transformation, i.e., the following identity holds:
\begin{eqnarray}
    \delta_{ij}\delta_{kl}
-
    \sum_{a=1}^3
    \left(\gamma_5\tau^a\right)_{ij}
    \left(\gamma_5\tau^a\right)_{kl}
&=&
    \Bs
    \begin{array}[t]{l}	\displaystyle
	\left(\Tate
	    \Omega(x)\Omega(x)
	\right)_{ij}
	\left(\Tate
	    \Omega(x)\Omega(x)
	\right)_{kl}
    \\\Tatetate	\displaystyle
    -
	\sum_{a=1}^3
	\left(\Tate
	    \Omega(x)
	    \gamma_5 \tau^a
	    \Omega(x)
	\right)_{ij}
	\left(\Tate
	    \Omega(x)
	    \gamma_5 \tau^a
	    \Omega(x)
	\right)_{kl}.
    \end{array}
\label{chiral.invariance.pi.sigma}
\end{eqnarray}
We need to establish  similar relations in  the  $qq$ sector.  For  this
purpose, it  is convenient  to  use  the  following Fierz identity  (for
details, see \Appendix{convensions.of.the.njl.model}.1.):
\begin{equation}
	{\cal L}_I
=
	\sum_\Gamma g_\Gamma (\Bar\psi\Gamma\psi)^2
=
	\sum_{\Gamma'} g^{(qq)}_{\Gamma'}
	(\Bar\psi {\Gamma'} \Bar\psi^T) (\psi^T \Gamma' \psi).
\end{equation}
In this representation, the identity \Eq{chiral.invariance} is expressed
in the following way:
\begin{equation}
    \sum_{\Gamma'} g^{(qq)}_{\Gamma'}
    \left(
	\Omega(x)
	\Gamma'
	\Omega(x)^T
    \right)_{ij}
    \left(
	\Omega(x)^T
	\Gamma'
	\Omega(x)
    \right)_{kl}
=
    \sum_{\Gamma'} g^{(qq)}_{\Gamma'} \Gamma'_{ij}\Gamma'_{kl}.
\label{chiral.invariance.diq}
\end{equation}
These $qq$ interactions also consist  of several closed chiral multiplet
sectors, and     in each sector   \Eq{chiral.invariance.diq}   is  valid
separately.  In particular, the $qq$  interaction in the scalar  diquark
channel ($J^{\pi} =  0^+$,  isoscalar, color  $\Bar{3}$) forms a  closed
chiral singlet, i.e.,
\begin{eqnarray}
\lefteqn{
    \sum_{A=2,5,7}
    \left(\Tate
	(\gamma_5 C^{-1})\tau_2 \beta_A
    \right)_{ij}
    \left(\Tate
	(C \gamma_5) \tau_2 \beta_A
    \right)_{kl}
}
\\
\nonumber
&=&
    \sum_{A=2,5,7}
    \left(\Tate
	\Omega(x)
	\left(\Tate
	    (\gamma_5 C^{-1})\tau_2 \beta_A
	\right)
	\Omega(x)^T
    \right)_{ij}
    \left(\Tate
	\Omega(x)^T
	\left(\Tate
	    (C \gamma_5) \tau_2 \beta_A
	\right)
	\Omega(x)
    \right)_{kl},
\end{eqnarray}
where  $\beta^A$  is  the rescaled color    Gell-Mann matrix  $\beta_A =
\sqrt{3\over          2}\lambda_A$   with     the          normalization
$\mbox{tr}(\beta_A\beta_B)  = 3\delta_{AB}$.   Note  that $\beta_A$  for
$A=2,5,7$ are anti-symmetric  matrices corresponding to the  color $\Bar
3_c$ diquark channels.  The $qq$ interaction in the axial vector diquark
($J^\pi  =  1^+$, isovector,  color  $\Bar{3}$) together with the vector
diquark ($J^\pi = 1^-$,   isoscalar, color  $\Bar{3}$) channel forms   a
closed chiral multiplet, i.e.,
\begin{eqnarray}
\lefteqn{
    \sum_{A=2,5,7}
    \Bs
    \begin{array}[t]{l}	\displaystyle
    \left[\;
	\sum_{a=1}^3
	\left(\Tate
	    (\gamma_{\mu} C^{-1}) (\tau_a\tau_2) \beta_A
        \right)_{ij}
        \left( \Tate
	    (C \gamma^{\mu}) (\tau_2\tau_a) \beta_A
        \right)_{kl}
    \right.
    \\\Tatetate\displaystyle
    \left.
    -	\left(\Tate
	    (\gamma_{\mu}\gamma_5 C^{-1}) \tau_2 \beta_A
	\right)_{ij}
	\left(\Tate
	    (C\gamma_5\gamma^{\mu}) \tau_2 \beta_A
	\right)
    \right]
    \end{array}
}
\\
\nonumber
&=&
    \sum_{A=2,5,7}
    \Bs
    \begin{array}[t]{l}	\displaystyle
    \left[\;
	\sum_{a=1}^3
	\left(\Tate
	    \Omega(x)
	    \left(\Tate
		(\gamma_{\mu} C^{-1}) (\tau_a\tau_2) \beta_A
	    \right)
	    \Omega(x)^T
        \right)_{ij}
        \left( \Tate
	    \Omega(x)^T
	    \left(\Tate
	        (C \gamma^{\mu}) (\tau_2\tau_a) \beta_A
	    \right)
	    \Omega(x)
        \right)_{kl}
    \right.
    \\\Tatetate\displaystyle
    \left.
    -	\left(\Tate
	    \Omega(x)
	    \left(\Tate
		(\gamma_{\mu}\gamma_5 C^{-1}) \tau_2 \beta_A
	    \right)
	    \Omega(x)^T
	\right)_{ij}
	\left(\Tate
	    \Omega(x)^T
	    \left(\Tate
		(C\gamma_5\gamma^{\mu}) \tau_2 \beta_A
	    \right)
	    \Omega(x)
	\right)_{kl}
    \right]
    \end{array}
\end{eqnarray}
We    give  a    list     of   these closed    chiral     multiplets  in
\Appendix{convensions.of.the.njl.model}.1.

Now we consider whether the sufficient condition \Eq{eq.24} is satisfied
in  the case  of  the ladder truncation   scheme,  i.e., the \threeqBS{}
kernel  of the  type  \Fig{fig3}.   In  this case,  the  2PI interaction
consists of  only  an  elementary  contact  interaction.   In  practical
numerical calculations, it is further truncated according to the quantum
numbers of  the diquark channels, i.e.,  the scalar diquark channel, the
axialvector  diquark channel, etc.  We  assume that the truncated vertex
corresponds to  a sum of closed  chiral multiplets. \Fig{fig5} shows the
steps involved in the analysis of the gauge transformation properties of
the kernel.
In the  first step, we  apply the local gauge transformation $\Omega(x)$
in order to get the transformed kernel  (l.h.s.of \Eq{eq.24}).  Only the
constituent  quark  propagators  transform, and  their transformation is
given by \Eq{gauge.transformation.propagator}.   In the  second step, we
use \Eq{chiral.invariance.diq},  and  the last step is  due  to the fact
that amputated   propagators are  delta functions.    As a   result, the
condition \Eq{eq.24}   is  satisfied  by    the  kernel in   the  ladder
approximation,  provided  that the  truncation of $qq$  channels is done
such as to have closed chiral multiplets.  In particular, since the $qq$
interaction  in the scalar diquark  channel  forms a chiral singlet, the
\threeqBS{} framework in  the ladder truncation  scheme keeping only the
$qq$ interaction  in the  scalar  diquark channel  gives rise  the  PCAC
relation correctly.  However, if the $qq$ interaction in the axialvector
diquark  channel is further  included, it  is  in principle necessary to
include also the  $qq$ interaction in   the vector diquark  channel.  In
practice, however, the  vector diquark channel can  be  expected to have
small  effects, at least  on the nucleon  mass, from the nonrelativistic
analogy.

Next we consider whether the  sufficient condition  is satisfied in  the
case of  the  kernel involving $q\Bar  q$ exchange, i.e.,  the \threeqBS{}
kernel of the type \Fig{fig4}.  \Fig{fig6} shows the argument for one of
the infinite   terms involved in  the  ladder sum.  We  assume that each
vertex corresponds to a  sum of closed  chiral multiplets.  In the first
step     we     apply      the       gauge       transformation    using
\Eq{gauge.transformation.propagator} to transform the constituent  quark
propagator.  In  the  second step  we  use \Eq{chiral.invariance}.  Note
that all  the phase factors, which  appear around the internal vertices,
cancel themselves.  The last step is due  to the fact that the amputated
propagator is just a delta function.   This demonstrates the validity of
\Eq{eq.24}   for  the  kernel involving   $q\bar{q}$  exchange.  Several
comments are in order.  First,  since the mass  of the pion is so small,
the $q\Bar{q}$ exchange in the  pionic channel will contribute much more
than the one in the sigma mesonic channel.  However, in principle, it is
necessary to include  both channels in   order to get the PCAC  relation
correctly.  Second, if we include the $q\Bar{q}$ exchange interaction in
the  \threeqBS{} kernel, it is necessary  to take into  account also the
``meson exchange   current  contributions'' in   the calculation  of the
matrix element of the  axial current, since  the l.h.s. of our criterion
\Eq{ward.id}   involves also the second    diagram in \Fig{fig4}.(b), as
discussed in Section 3.

These examples    are   straightforwardly  extended to   more   general,
complicated and realistic cases (for example, to the expansion scheme of
\cite{lorenz1}).  We may consider  these \threeqBS{} kernel as  a formal
sum of 2PI and 3PI diagrams.  It is thus  not hard to convince ourselves
that the chiral symmetry in the original interaction Lagrangian leads to
the PCAC relation correctly in our formalism, provided the truncation of
the kernel is done consistent with the condition (\ref{eq.24}).

\section{The Goldberger-Treiman Relation and $g_{\pi NN}$}

In this section, we give the explicit numerical results for $g_{\pi NN}$
in the chiral limit  together with the   off-shell $g_{\pi NN}$  off the
chiral limit.  We restrict our attention to the ladder truncation scheme
keeping   only the  $qq$ interaction    in  the scalar diquark  channel.
Although this truncation scheme of \threeqBS{} kernel should lead to the
correct   Goldberger-Treiman  relation, the   UV    regularization  (the
Euclidean    cutoff in  our  case)  unfortunately   violates the  chiral
symmetry.  One of the aims of this  section is therefore to estimate the
degree of violation of the  PCAC/Goldberger-Treiman relation due to  the
UV-regularization.  In this section, unless explicitly indicated for the
more general case,   {\em the  external   fields are understood  to   be
absent}, i.e., $v=a=0$,  $m=m_0$. For consistency  reason, we  prefer to
use the axial current operator $A^b_{\mu}(x)$  as an interpolating field
for  the  pion  in  this paper.   For those  readers  who prefer  to use
$\displaystyle   \Bar\psi(x)   \gamma_5  {\tau^b\over  2}\psi(x)$  as an
interpolating field   for  the pion,  the  following   relation would be
convenient\footnote{We  will not use this   relation explicitly in  this
paper.}:
\begin{equation}
    A_{\mu}^b(x)
=
    f_{\pi} \partial_{\mu} \pi^b_{\rm as}(x) + \cdots,
\end{equation}
where $\pi^b_{\rm   as}(x)$ is the  asymptotic field   operator for the
pion.

\subsection{The Extraction of $g_{\pi NN}$ in the Chiral Limit}
\label{section.5.1}

In this subsection,  $m_0  = 0$ is  understood.  The  axial form factors
$g_A(q^2)$ and $h_A(q^2)$ are defined through the following relation:
\begin{equation}
    \Bar{u}^{(\alpha)a}(\vec{P})
    {\tau^b\over 2}
    \left( \Tate
	g_A(q^2)\gamma^{\mu}\gamma_5
    +	q^{\mu}h_A(q^2)\gamma_5
    \right)
    u^{(\beta)c}(\vec{L})
=
    \left\langle N^{(\alpha)a}(\vec{P}) \left| \Tate
	A^{b\mu}(x=0)
    \right| N^{(\beta)c}(\vec{L}) \right\rangle,
\label{goldberger.1st.relation}
\end{equation}
where $\Bar{u}^{(\alpha)a}(\vec{P})$ and   $\Bar{u}^{(\beta)c}(\vec{L})$
are the final and initial nucleon bispinors with the isospins $a,c$, the
helicities  $\alpha,\beta$ and the momenta  $P,L$ ($P^2 = L^2 = m_N^2$),
respectively. $q = P -  L$ is the momentum  transfer.  $g_A$ and $g_{\pi
NN}$ are extracted as follows\cite{itzykson}:
\begin{equation}
    g_A
=
    g_A(q^2 = 0),
\Hs
    h_A(q^2)
=
    {-2f_\pi g_{\pi NN} \over q^2}
+
    \cdots.
\end{equation}
A      multiplication     of     $q_{\mu}$    on     both       sides of
\Eq{goldberger.1st.relation} leads to
\begin{eqnarray}
	0
&=&
	\Bar{u}^{(\alpha)a}(\vec{P})
	{\tau^b\over 2}
	\left( \Tate
		g_A(q^2)\fslash{q}\gamma_5
	+	q^2h_A(q^2)\gamma_5
	\right)
	u^{(\beta)c}(\vec{L})
\\
\nonumber
&=&
	\Bar{u}^{(\alpha)a}(\vec{P})
	{\tau^b\over 2}
	\left( \Tate
		2 m_N g_A(q^2)\gamma_5
	+	q^2 h_A(q^2)\gamma_5
	\right)
	u^{(\beta)c}(\vec{L})
\\
    0
&=&
    2 m_N g_A(q^2) + q^2 h_A(q^2),
\label{pre.goldberger.treiman.relation}
\end{eqnarray}
where we used the PCAC relation  (\ref{desired.relation}) for $m_0 = 0$.
As is  well-known\cite{itzykson}, due to the  pion pole of $h_A(q^2)$ at
$q^2 = 0$, this relation leads to the Goldberger--Treiman(GT) relation:
\begin{equation}
	m_N g_A = f_{\pi} g_{\pi NN}.
\end{equation}

We expand the both sides of \Eq{goldberger.1st.relation} with respect to
a small $q^2$   to extract $g_{\pi  NN}$.   The  l.h.s. is  expanded  as
follows:
\begin{eqnarray}
\lefteqn{
    \Bar u^{(\alpha)a}(\vec P)
    {\tau^b\over 2}
    \left(\Tate
	g_A(q^2) \gamma^{\mu}\gamma_5 + q^{\mu}h_A(q^2) \gamma_5
    \right)
    u^{(\beta)c}(\Lambda^{-1}\vec P)
}
\\\nonumber
&=&
    \Bar u^{(\alpha)a}(\vec P)
    {\tau^b\over 2}
    \left(\Tate
	g_A(q^2) \gamma^{\mu}\gamma_5 + q^{\mu}h_A(q^2) \gamma_5
    \right)
    \hat S(\Lambda^{-1})
    u^{(\beta)c}(\vec P)
\\\nonumber
&\simeq&
    \left(\Tate
        \Bar u^{(\alpha)a}(\vec P)
	{\tau^b\over 2} \gamma^{\mu}\gamma_5
        u^{(\beta)b}(\vec P)
    \right)
    g_{A}
+
    \left(\Tate
	\Bar u^{(\alpha)a}(\vec P)
	{\tau^b\over 2} \gamma_5
	S(-\lambda)
	u^{(\beta)c}(\vec P)
    \right)
    q^{\mu}
    {-2 f_{\pi} g_{\pi NN} \over q^2}
+
    O(q),
\end{eqnarray}
where ${(\Lambda^{-1})_{\mu}}^{\nu} = {(e^{-\lambda})_{\mu}}^{\nu}$ is a
boost   matrix, i.e.,  $L_{\mu} =  {(\Lambda^{-1})_{\mu}}^{\nu}P_{\nu}$,
$\displaystyle \hat  S(\Lambda^{-1}) =  e^{S(-\lambda)}$  is the   boost
matrix in  the Dirac  bispinor space  with $\displaystyle S(-\lambda)  =
(-\lambda)_{\mu\nu}{-i\over  4} \sigma^{\mu\nu}$.   Note that $q_{\mu} =
P_{\mu}      -       {(\Lambda^{-1})_{\mu}}^{\nu}P^{\nu}          \simeq
{\lambda_{\mu}}^{\nu}P_{\nu} + O(\lambda^2)$.  We  thus have $\lambda  =
O(q)$.  Hence we have
\begin{equation}
    \left(\Tate
	\Bar u^{(\alpha)a}(\vec P)
	{\tau^b\over  2} \gamma_5  S(-\lambda)
	u^{(\beta)c}(\vec  P)
    \right)
    q^{\mu}
=
    O(q^2),
\end{equation}
which cancels the pion pole of $h_A(q^2)$ at $q^2 = 0$.

The   r.h.s.    of \Eq{goldberger.1st.relation}   is   evaluated  by the
\threeqBS{}      /Faddeev      expression     \Eq{matrix.element.1}  and
\Eq{matrix.element.faddeev.equation} of the nucleon matrix element:
\begin{eqnarray}
    \left\langle N^{(\alpha)a}(\vec P) \left| \Tate
	A^{b\mu}(x=0)
    \right| N^{(\beta)c}(\vec L) \right\rangle
\label{gpinn.faddeev}
&=&
    \tildebar{{\Psi}}
    [N^a_{\alpha}(\vec P)]
    \left[
	{i\delta K^{[e]} \over \delta a_{\mu}^b(x=0) }
    \right]^{[0]}
    \Psi[N^c_\beta(\vec L)]
\\\nonumber
&=&
    \Bar\psi[N^a_{\alpha}(\vec P)]
    \left[
	{i\delta K^{[e]}_F \over \delta a_{\mu}^b(x=0)}
    \right]^{[0]}
    \psi[N^c_{\beta}(\vec L)],
\end{eqnarray}
where $\psi[N^c_{\beta}(\vec  L)]$,  $\Bar\psi[N^a_\alpha(\vec P)]$  and
$K^{[e]}_F$    are  the Faddeev    amplitudes  and  the Faddeev  kernel,
respectively.        These       quantities        are   defined      in
\Appendix{appendix.faddeev.equation}, where  also their relation  to the
\threeqBS{} quantities is explained. The Faddeev equation is depicted in
\Fig{fig.faddeev}.    Since we adopted the   ladder truncation scheme in
this section, the $qq$ interaction in \Fig{fig.faddeev} is understood to
be point-like and separable.  Hence the double line in \Fig{fig.faddeev}
can be expressed as a  geometric series of   $qq$ bubble integrals.   In
addition, since the $qq$ interaction  is truncated to the scalar diquark
channel,   the double line    in \Fig{fig.faddeev}  now  stands for  the
following t-matrix $t_{\rm sd}(q^2)$ in the scalar diquark channel:
\begin{equation}
    t_{\rm sd}(q^2)
=
    2\times {2 i g_{\rm sd}\over 1 - 2 i g_{\rm sd} \Pi_{\rm sd}(q^2)},
\end{equation}
with
\begin{equation}
    \Pi_{\rm sd}(q^2)\delta_{AB}
=
    \int {d^4k\over(2\pi)^4}
    \mbox{tr}
    \left(\Tate
	(C\gamma_5 \tau_2 \beta_A)
	iS_F(q + k)
	(\gamma_5 C^{-1} \tau_2 \beta_B)
	iS_F(-k)^T
    \right).
\end{equation}
Since  the second  line in  \Eq{gpinn.faddeev}  involves  the functional
derivative  of the Faddeev kernel,   the contributions are classified as
the following three   types\footnote{In the diagrams  of  \Fig{fig7}, we
adopted the  relative momenta  of the spectator   quark and the diquark,
which are defined in  \Eq{relative.total.momentum}, with the value $\eta
=    1/2$.}(cf.   \Appendix{appendix.a.3}):   (a)   the    quark current
contribution[\Fig{fig7}.(a)],     (b)       the      exchange    current
contribution[\Fig{fig7}.(b)],    (c)      the      diquark       current
contribution[\Fig{fig7}.(c)].  Note that,  due to  the iso-scalar nature
of the  scalar diquark, the  diquark current  contribution to the matrix
element of the iso-vector axial current vanishes identically.

The evaluation  of the  diagram \Fig{fig7}.(a)  posses  a problem, since
there appears a  delta function $\delta^{(4)}(p  - l - q/2)$ in addition
to the delta  function associated with  the total momentum conservation.
Therefore,  depending  on the  momentum  transfer   $q$, the  evaluation
involves the values of the Faddeev amplitude at points which are outside
the  mesh  used for  the solution  of  the Faddeev equation\footnote{The
numerical procedure  to   obtain the mass  and   the  associated Faddeev
amplitudes is explained in  detail in \cite{ishii1,ishii2,ishii3}.}.  To
avoid    this  problem,  we    use  the   homogeneous  Faddeev  equation
\Eq{homogeneous.faddeev.eq}[\Fig{fig.faddeev}] to  iterate   the Faddeev
amplitude in  the  final  state.   The diagram  \Fig{fig7}.(a)   is thus
exactly equivalent to \Fig{fig7}.(a'), which is free from the additional
delta  function.   We   therefore  have  to  evaluate   the  diagrams of
\Fig{fig7}.(a') and \Fig{fig7}.(b).   Since the operator  insertion on a
constituent     quark      line    has    been       given     in   Eqs.
(\ref{functional.derivative.propagator})    --       (\ref{eq.37})   and
\Fig{fig2}.(a), we are led to the four diagrams shown in \Fig{fig7}.(d).
Here, we  used  the  identity  \Eq{matrix.element.rest.frame.faddeev} to
express the matrix  element only in terms of  the Faddeev amplitudes  in
the rest frame, and used also the identity
\begin{equation}
	\gamma^{\mu}(\Lambda p)_{\mu}
=
	S(\Lambda)  \fslash{p} S(\Lambda^{-1}).
\label{eq.sps}
\end{equation}
Note that, since we truncated the $qq$ interaction to the scalar diquark
channel,  two of the  boost matrices $\hat   S(\Lambda^{-1})$ out of the
three      ($\hat              S(\Lambda^{-1})^{\otimes      3}$)     in
\Eq{matrix.element.rest.frame.faddeev}  cancel.  The  remaining    $\hat
S(\Lambda^{-1})$  is  found in  \Fig{fig7}.(d)  at  the point where  the
operator   is inserted,   which is  a   consequence  of the manipulation
\Eq{eq.sps}.  By using the  following relations, which express the first
order deviations due to the non-vanishing momentum transfer $q$:
\begin{eqnarray}
    \delta \left(\Tate S(\Lambda^{-1}) \right)
&=&
    S(-\lambda)
\\\nonumber
    \delta\left(\Tate \gamma^{\mu}(\Lambda p)_{\mu} \right)
&=&
    \left[\Tate S(\lambda), \fslash{p} \right]
\\\nonumber
    \delta\left(\Tate S_F(P - \Lambda^{-1}(P/2 - l)) \right)
&=&
    S_F(P/2 + l)
    \left[\Tate S(-\lambda),\fslash{P}/2 - \fslash{l} \right]
    S_F(P/2 + l),
\Hs
    \mbox{etc.},
\end{eqnarray}
we can consider   the  limit $q \to   0$.  After  using  the homogeneous
Faddeev equation \Eq{homogeneous.faddeev.eq}, we are left with the eight
diagrams depicted in \Fig{fig8}.  Note that the equality $\Fig{fig7}.(a)
+ \Fig{fig7}.(b) \left(\Tate = \Fig{fig7}.({\rm d})\right) = \Fig{fig8}$
holds only in the limit $q \to 0$, which  is emphasized in \Fig{fig8} by
``$q \to 0$''.  By comparing  both sides of \Eq{goldberger.1st.relation}
in the limit $q \to 0$, the explicit spin-parity projection\footnote{The
spin-parity projection  is performed by   using the helicity  formalism.
(See  for  detail\cite{ishii3})} shows that   the first two diagrams are
proportional  to $\displaystyle \Bar  u^{(\alpha)a}(\vec P) {\tau^b\over
2}  \gamma^{\mu}\gamma_5  u^{(\beta)b}(\vec  P)$,  which  contribute  to
$g_A$, and  the other diagrams  are proportional to  $\displaystyle \Bar
u^{(\alpha)a}(\vec     P)     {\tau^b\over   2}    \gamma_5  S(-\lambda)
u^{(\beta)b}(\vec P)$, which contribute to $g_{\pi  NN}$.  Note that the
1st, the 3rd,the  4th and the 5th diagrams  in \Fig{fig8}  come from the
quark current contribution[\Fig{fig7}.(a)], and the others come from the
exchange current contribution[\Fig{fig7}.(b)].

\subsection{Numerical Results in the Chiral Limit}

We first have  to explain the choice  of the parameters.  There are four
parameters:  the cutoff $\Lambda$ (Euclidean  sharp cutoff), the current
quark mass $m_0$, the effective coupling  constant in the pionic channel
$g_{\pi}$ and the effective coupling constant in the $qq$ scalar diquark
channel $g_{\rm sd}$.  We fix the first three parameters ($\Lambda, m_0,
g_{\pi}$) by solving the gap equation \Eq{explicit.gap.equation} for the
constituent quark  mass $M$,  and  \Eq{vacuum.parameter.1} for  the pion
mass $m_{\pi}$  and  decay  constant $f_{\pi}$, imposing  the  following
three conditions: (1) $m_{\pi}= 140$ MeV, (2) $f_{\pi} = 93$ MeV, (3) $M
=  400$ MeV.  The resulting  values are $\Lambda  = 739$ MeV, $g_{\pi} =
10.42$ GeV$^{-2}$, $m_0  = 8.99$ MeV.  Once  these parameters are fixed,
we  consider the chiral limit by  taking  $m_0 \to 0$ keeping $\Lambda$,
$g_{\pi}$ fixed.      \Eq{explicit.gap.equation}   provides the    $m_0$
dependences of $M = M(m_0)$.  We treat  $g_{\rm sd}$ as a free parameter
(independent of $m_0$) to generate different nucleon masses.  Note that,
in the case $m_{\pi} = 140$  MeV, $g_{\rm sd}/g_{\pi}  = 0.66$ gives the
experimental value   of  the  nucleon  mass $m_N    =  940$  MeV.    For
convenience, we  plot the nucleon mass $m_N$  (for $g_{\rm sd}/g_{\pi} =
0.66$), the quark-diquark  threshold($m_N  + m_{\rm sd}$) where  $m_{\rm
sd}$ is the scalar diquark mass, the pion mass $m_{\pi}$, the pion decay
constant $f_{\pi}$,  and  the constituent  quark  mass $M$   against the
current quark mass $m_0$ in \Fig{FigN1}.  We  also plot the nucleon mass
in the chiral limit against $g_{\rm  sd}/g_{\pi}$ [solid line], together
with   the   physical  case($m_{\pi}  =   140$ MeV)    [dashed  line] in
\Fig{FigN2}.

Our results  for $g_{\pi NN}$ and $g_A$  in the chiral limit are plotted
against   $g_{\rm  sd}/g_{\pi}$ by the   solid   line in \Fig{FigN3} and
\Fig{FigN4},  respectively.  We obtain $g_{\pi  NN} = 13.2$ and $g_{A} =
1.32$ for  $g_{\rm sd}/g_{\pi} =   0.66$  compared to  the  experimental
values $g_{\pi NN}^{\rm (exp)} =  13.5$ and $g_A^{\rm(exp)} = 1.26$.  We
will discuss the extension off the chiral limit in the next subsection.

To estimate the violation of the GT relation, it is convenient to define
a quantity:
\begin{equation}
    \Delta_{\rm G}
\equiv
    {f_{\pi} g_{\pi NN} \over m_N g_A}.
\end{equation}
In the chiral  limit, the  deviation of   $\Delta_{\rm G}$  from $1$  is
solely  due to the cutoff   artifact.  We plot  $\Delta_{\rm G}$ against
$g_{\rm sd}/g_{\pi}$ in \Fig{FigN5} in   the chiral limit [solid  line].
It is seen  that the violation of  the GT relation  is up to  4  \%.  In
particular,  for the reasonable  case  $g_{\rm sd}/g_{\pi} = 0.66$,  the
violation is only 2 \% in the chiral limit.

\subsection{PCAC Violation off the Chiral Limit}

In  the  chiral  limit, $\Delta_{\rm G}$  works   as the measure of  the
violation of the PCAC relation due to the cutoff artifact.  However, off
the chiral  limit,  $\Delta_{\rm G}$  does not work  as the  measure any
more, since it  contains, in addition  to the unphysical cutoff artifact
which we  are  going  to estimate  here,   the physical  effect  of  the
non-vanishing current quark mass.   Therefore, in order to  estimate the
cutoff artifact, we need to construct a quantity which picks up only the
cutoff artifact for any values of $m_0 > 0$.   To this end, we introduce
another form factor $i_A(q^2)$ through the following relation:
\begin{equation}
    \left\langle N^a_{\alpha}(\vec P) \left| \Tate
	i \Bar\psi(x) \gamma_5 {\tau^b\over 2} \psi(x)
    \right| N^c_{\beta}(\vec L) \right\rangle
=
    e^{iqx}
    \Bar u^a_{\alpha}(\vec P)
    \left( i \gamma_5 {\tau^b\over 2} \right)
    u^c_{\beta}(\vec L)
    \times
    i_A(q^2).
\end{equation}
\Eq{pre.goldberger.treiman.relation} generalizes in the following way:
\begin{equation}
    2m_N g_A(q^2) + q^2 h_A(q^2) = 2m_0 i_A(q^2).
\end{equation}
In particular, in the limit $q^2 \to 0$,  since now $m_{\pi} \neq 0$, we
have
\begin{equation}
    2m_N g_A = 2m_0 i_A(q^2 = 0).
\end{equation}
Now we define a quantity $\Delta_{\rm P}$ as follows:
\begin{equation}
    \Delta_{\rm P}
=
    {m_0 i_A(q^2 = 0) \over m_{N} g_A}.
\end{equation}
By construction, the  deviation of $\Delta_{\rm P}$ from  1 is solely due
to the cutoff artifact for any value of  $m_0$.  To evaluate $i_A(q^2)$,
we first have to solve the following equation for the operator insertion
on       a      constituent        quark         line   similar       to
\Eq{functional.derivative.propagator}:
\begin{eqnarray}
    \left[
	{\delta S_F^{[e]}(x,y) \over \delta p^b(z)}
    \right]\Zero
&=&
    \Bs
    \begin{array}[t]{l}	\displaystyle
	S_F\Zero(x,z)
	\left(\Tate i \gamma_5 {\tau^b \over 2} \right)
	S_F\Zero(z,y)
    \\	\displaystyle
    +	
	\int d^4 z'
	S_F\Zero(x,z')
	\left[
	    {\delta \Sigma^{[e]}(z')\over \delta p^b(z)}
        \right]\Zero
	S_F\Zero(z',y).
    \end{array}
\label{functional.derivative.propagator.pb}
\\\nonumber
    \left[
    {
    	\delta \Sigma^{[e]}(z')
    \over
	\delta p^b(z)
    }
    \right]
    \Zero
&=&
    \sum_\Gamma 2ig_{\Gamma}^{(q\Bar q)}
    \Tr\left(
	\left[
	{
	    \delta S^{[e]}_F(z',z')
	    \over
		\delta p^b(z)
	}
	\right]
	\Zero
	\Gamma
    \right)
    \Gamma.
\end{eqnarray}
Since $g_{\rm ax}=0$, we can parameterize $\delta \Sigma/\delta p^b$ as
\begin{equation}
    \int d^4 z' e^{iq(z'-z)}
    \left[
	{\delta \Sigma^{[e]}(z') \over \delta p^b(z)}
    \right]
=
    I(q^2) \left(\Tate i\gamma_5 {\tau^b \over 2} \right).
\end{equation}
$I(q^2)$ then satisfies the following equation:
\begin{equation}
    I(q^2)
=
    -2ig_{\pi} \Pi_{55}(q^2)
-
    2ig_{\pi} \Pi_{55}(q^2) I(q^2),
\end{equation}
with the solution
\begin{equation}
    I(q^2)
=
    {-2ig_{\pi} \Pi_{55}(q^2) \over 1 + 2ig_{\pi} \Pi_{55}(q^2)}.
\end{equation}
Note that this quantity of course has the pion  pole.  The evaluation of
$i_A(q^2  = 0)$   amounts  to the  2nd   to  the 8th  diagrams given  in
\Fig{fig8},  where $q_{\mu}\tilde H(q^2)$ is  replaced by  $1 + I(q^2)$.
In \Fig{FigN6}, we plot $\Delta_{\rm  P}$ against the current quark mass
$m_0$ for $g_{\rm sd}/g_{\pi}  = 0.66$ case [dashed  line].  It  is seen
that the cutoff artifact is again within 4\%.

\subsection{The Off-shell $g_{\pi NN}$ off the Chiral Limit}
\label{section.off.shell.gpinn.off.the.chiral.limit}

We next  evaluate  $g_{\pi NN}$ off the   chiral limit ($m_{\pi} =  140$
MeV).  In this case, whereas the definition of  $g_{A} = g_{A}(q^2 = 0)$
does not  change\footnote{The evaluation  of the iso-vector  $g_{A}$ off
the chiral limit  was already done in ref.  \cite{asami} in the case  of
$g_{\rm  ax} = 0$.},   since the pion   pole  of $h_A(q^2)$  is shifted,
$g_{\pi NN}$ is extracted according to:
\begin{equation}
    h_A(q^2)
=
    {-2f_{\pi}g_{\pi NN} \over q^2 - m_{\pi}^2}
+   \cdots.
\label{extract.g.pi.nn}
\end{equation}
In order to extract $g_{\pi NN}$ from the nucleon  matrix element of the
axial current,  it is necessary to evaluate  the form factor $h_A(q^2)$.
However, at this stage it is still difficult  to evaluate $h_A(q^2)$ for
non-vanishing     momentum  transfer    in  the  relativistic    Faddeev
approach\footnote{The numerical evaluation  of the on-shell $g_{\pi NN}$
off  the chiral limit  is  currently under consideration.}.  We  confine
ourselves to  evaluate the off-shell $g_{\pi NN}(q^2  = 0) \equiv \tilde
g_{\pi NN}$ defined    by  the single-pole dominance  approximation   to
$h_A(q^2)$ as
\begin{equation}
    \tilde g_{\pi NN} \equiv {m_{\pi}^2 \over 2 f_{\pi}} h_A(q^2 = 0).
\label{eq.66}
\end{equation}
Note  that, whereas the  nearest cut in  the physical  $h_A(q^2)$ is the
three pion cut ($q^2 >  9 m_{\pi}^2$), the  Cutokosky rule(cf.  p.315 in
\cite{itzykson}) suggests that the nearest  cut in our $h_A(q^2)$ is the
$q\Bar q$ cut  ($q^2 > 4M^2$).  This  is mainly  due  to the  mean field
approximation  for   the  vacuum, and  due   to also  the   leak  of the
confinement in the NJL model.  Since $4M^2$ is larger than $9m_{\pi}^2$,
the single-pole dominance approximation may work better in our case.  We
note that, although  we could subtract the  $q\Bar q$ cut  contributions
from $\tilde H(q^2)$, it is impossible to subtract it from the remaining
part, because   the calculation  refers only    to $q=0$.  The  explicit
evaluation shows that the subtraction of the $q\Bar q$ cut contributions
from $\tilde H(q^2)$ leads to only a small difference.  We note that the
quantity $\tilde g_{\pi  NN}$ is not only one  of the possible off-shell
extensions of the on-shell $g_{\pi NN}$, but also the value of the axial
form factor $h_A(q^2)$  at $q^2  =  0$ up to the  well-defined numerical
factor given by  \Eq{eq.66}.   This enables  us to derive  an analytical
expression of $\Delta_{\rm G}$ which  follows from the PCAC relation  by
neglecting      the        small      cutoff    artifact.           (See
\Appendix{appendix.non-vanishing.gax.gtv.index}.)

We plot $\tilde g_{\pi  NN}$ against $g_{\rm sd}/g_{\pi}$ in \Fig{FigN3}
[dashed line],  and $g_{A}$ in \Fig{FigN4} [dashed  line].  The value of
$\tilde g_{\pi NN}$ is 13.5 for  $g_{\rm sd}/g_{\pi} = 0.66$. This value
is  quite reasonable compared to the  experimental value.   We also plot
$\Delta_{\rm G}$ off the chiral limit [dotted line] and $\Delta_{\rm P}$
[dashed line]  against $g_{\rm sd}/g_{\pi}$  in \Fig{FigN5}.  The reader
might suspect why  the validity of the  GT relation could be improved by
going  off  the  chiral  limit.  The   reason is   that  the effect   of
non-vanishing $m_0$ works  into the opposite  direction compared  to the
UV-cutoff artifact.   To see this, we plot  $\Delta_{\rm G}$ against the
current   quark  mass $m_0$  [solid  line]  in  \Fig{FigN6} for the case
$g_{\rm sd}/g_{\pi}  = 0.66$.   We  also  plot $\Delta_{\rm P}$  against
$m_0$ [dashed line],  which is used to indicate  the size of  the cutoff
artifact  contained in $\Delta_{\rm G}$.  It   is seen that $\Delta_{\rm
G}$ is  a  monotonically  decreasing function   of $m_0$.   However, the
experimental data $\Delta_{\rm  G} = 1.06$  suggests that, as far  as we
believe that  $m_{\pi} = 140$ MeV is  really close to  the chiral limit,
$\Delta_{\rm  G}$ should be an   increasing function in the vicinity  of
$m_0   \simeq    0$.     We  investigate  this     problem   further  in
\Appendix{appendix.non-vanishing.gax.gtv.index}  by  taking into account
also the   effects of non-vanishing coupling   constant in the isovector
axial  vector $q\bar{q}$ channel $g_{\rm ax}$.   (We leave this analysis
for the  appendix, because the non-vanishing  $g_{\rm  ax}$ makes things
quite complicated.)
The main conclusions there  are summarized as follows:
(1) $g_A$ and $\tilde g_{\pi  NN}$ both decrease with increasing $g_{\rm
ax}/g_{\pi}$, and increase with increasing $m_0$.
(2) The best fit of $g_{A}$ and $\tilde g_{\pi NN}$ could be obtained in
the  region $0  \le g_{\rm  ax}/g_{\pi}  <  0.1$, which, however,  would
depend on the quantity  which we prefer to  adjust.  From this point  of
view, $g_{\rm ax} = 0$ is actually a rather good choice, because $\tilde
g_{\pi NN}$ is very reasonable and $g_{A} =  1.33$ is still close to the
experimental value $g_A^{\rm (exp)} = 1.26$.
(3) For those values of $g_{\rm ax}$ which we examine, $\Delta_{\rm G}$
remains to be a decreasing function of $m_0$.
(4)     An    analytic       expression      of     $\Delta_{\rm     G}$
(\Eq{gtv.mesonic.expression}) is derived by neglecting the small cut-off
artifact and  by  assuming  that   the vacuum  is  approximated  by  the
mean-field approximation.   All  the baryonic quantities  disappear from
this expression.  In  particular,  this expression is valid  even beyond
the ladder  truncation scheme   for  the 3qBS kernel.    The discrepancy
between the analytic  and the  numeric $\Delta_{\rm  G}$  is due to  the
cutoff artifact. It is found to be within 3 \%.
(5)
As a consequence, to make $\Delta_{\rm  G}$ to be an increasing function
of $m_0$ and to obtain $\Delta_{\rm G} = 1.06$, we have to go beyond the
validity  of this analytic  expression  of $\Delta_{\rm G}$.  Therefore,
all we  can do is  either  to improve the   gap equation for  the vacuum
beyond the mean field approximation or to  estimate the on-shell $g_{\pi
NN}$.
We do not further investigate this problem in this paper.

\section{Summary and Discussions}

In this  work we reviewed how to  evaluate expectation values of quantum
one-body operators in the framework of the \threeqBS{} /Faddeev equation
by introducing  classical external fields as a  technical  tool.  In the
\threeqBS{}   approach,  the  expectation    values   are   obtained  by
sandwiching, the  functional  derivative of the  \threeqBS{} kernel with
respect to  the  corresponding  external field between   the \threeqBS{}
amplitudes.   In  the  Faddeev  approach,  the  expectation  values  are
obtained by sandwiching,  between the Faddeev amplitudes, the functional
derivative of    the Faddeev kernel  with  respect  to the corresponding
external  field.  We gave the  criterion for \threeqBS{} kernels to give
rise  to the  PCAC relation  correctly.  For practical  purpose, we also
gave the  sufficient  condition for  the validity of  this  criterion by
introducing  the local  ``axial'' gauge  transformation  of the external
fields.   We  applied the  sufficient   condition to several \threeqBS{}
kernels.  The main results are as follows: (1) If the \threeqBS{} kernel
is  truncated  in the ladder  truncation  scheme  keeping  only the $qq$
interaction in the scalar diquark channel, the PCAC relation is obtained
correctly.  (2)   If the $qq$ interaction  in  the  axial-vector diquark
channel is   included, it  is   necessary   to include  also  the   $qq$
interaction  in the  vector  diquark channel to   give rise to  the PCAC
relation correctly.  (3)   Even  if the  $qq$  interaction due   to  the
$q\Bar{q}$ exchange in both the pionic and  the sigma mesonic channel is
included, the correct  PCAC relation is  obtained.  Concerning the point
(2), we  note that the vector diquark  channel is often considered to be
not  important.    This is   because the  non-relativistic   quark model
suggests that the contribution from this channel to  the nucleon mass is
suppressed   in the non-relativistic   limit.   However, to respect  the
chiral $SU(2)_L\times SU(2)_R$   symmetry, the $qq$ interaction in   the
vector  diquark channel should  be included,  even  if it is expected to
give a negligible   contribution to the nucleon  mass.   We should note,
however, that  the relativistic Faddeev  equation including all the $qq$
interactions in the  scalar,   axial-vector and vector  diquark  channel
amounts to a two-dimensional integral equation with $14\times 14$ matrix
structure  even after the  spin-parity projection  is carried out, which
requires a tremendous effort to be solved.

Although  these truncation  schemes   give  rise  to the PCAC   relation
correctly,  the  regularization scheme,    which   has been adopted   in
practical numerical calculations   so far, does  not  respect the chiral
symmetry, leading to the violation  of the Goldberger-Treiman  relation.
To estimate this violation, we carried out  the numerical evaluations of
$g_{A}$ and $g_{\pi NN}$ in the chiral limit in the simplest case, i.e.,
the  ladder truncation scheme keeping  only the $qq$  interaction in the
scalar diquark channel.  We  found that the violations  are up to  4 \%.
In particular, for the case $g_{\rm s}/g_{\pi} = 0.66$, which reproduces
the experimental nucleon mass, the violation is only 2  \%. The value of
$g_{\pi NN}$ in the chiral limit  is $13.2$ which  is quite close to the
experimental  value $13.5$, and  $g_{A}$ becomes $1.32$  compared to the
experimental value $1.26$.  We next estimated the  PCAC violation due to
the cutoff artifact off the chiral limit.   We found that this violation
is again within 4\%.

In the  relativistic Faddeev method,  it is still difficult to calculate
form  factors for non-vanishing  momentum  transfer, which is needed  to
extract the on-shell $g_{\pi  NN}$ off the  chiral limit.  So we defined
the off-shell $\tilde g_{\pi NN}$ by means  of the single-pole dominance
approximation to $h_A(q^2)$.    Although we obtained   a very reasonable
result $\tilde g_{\pi  NN} =  13.5$  for the case $g_{\rm  sd}/g_{\pi} =
0.66$, the effect  of  non-vanishing current  quark  mass $m_0$  on  the
Goldberger-Treiman violation ($\Delta_{\rm  G}(m_0)$) was found to be in
the ``wrong'' direction:  Whereas the experimental value of $\Delta_{\rm
G}$  is  $1.06$,  which suggests that   $\Delta_{\rm G}$   should be  an
increasing  function  of $m_0$   in the   vicinity  of  $m_0 =  0$,  our
$\Delta_{\rm G}(m_0)$  is a decreasing function of  $m_0$.  We  tried to
resolve  this problem  (\Appendix{appendix.non-vanishing.gax.gtv.index})
by taking into  account  the effect of non-vanishing  effective coupling
constant in the iso-vector axial-vector mesonic channel $g_{\rm ax}$.
The  main results  are  as   follows:
(1) $g_A$ and $\tilde g_{\pi  NN}$ both decrease with increasing $g_{\rm
ax}/g_{\pi}$, and increase with increasing $m_0$.
(2) The best fit  of $g_A$ and $\tilde g_{\pi  NN}$ could be obtained in
the region $0   \le g_{\rm  ax}/g_{\pi}  < 0.1$,  which, however,  would
depend on the quantity which  we prefer to adjust.   From this point  of
view, $g_{\rm  ax} = 0$ is actually  a rather good choice, since $\tilde
g_{\pi NN} = 13.5$ is a very reasonable result and $g_A = 1.33$ is still
close to the experimental value $g_A^{\rm (exp)} = 1.26$.
(3) For those values of $g_{\rm ax}$ which we examined, $\Delta_{\rm G}$
remains to be a decreasing function of $m_0$.
(4) An analytic expression of $\Delta_{\rm G}$ was derived by neglecting
the   small   cutoff  artifact  and by    assuming  that  the vacuum  is
approximated   by the  mean-field method.  All   the baryonic quantities
disappear from this expression.  In particular, this expression is valid
even beyond the   ladder  truncation scheme for the    3qBS kernel.  The
discrepancy between the analytic and the numeric $\Delta_{\rm G}$ is due
to the cutoff artifact. It was found to be within 3 \%.
(5) As   a consequence, to make  $\Delta_{\rm  G}$ to  be  an increasing
function of  $m_0$, we have to take  into account  the effects which are
beyond the validity of the analytic expression of $\Delta_{\rm G}$.
Hence, all  we  can do   are either to   improve  the vacuum beyond  the
meanfield approximation or to estimate the on-shell $g_{\pi NN}$.

We finally  give  a  comment   on the  iso-scalar  $g_A^{(0)}$.  It   is
straightforward to  extend  our formalism  to the  chiral $U(1)_L \times
U(1)_R$ case.  The $U_A(1)$ anomaly in QCD is simulated in the NJL model
as an  explicit $U_A(1)$  symmetry breaking.   It  is easy to  see that,
whereas the vector, axial-vector and tensor diquark channels form closed
chiral $U(1)_L\times U(1)_R$ singlets separately,  only a combination of
the  scalar diquark and the pseudo  scalar diquark forms a closed chiral
$U(1)_L\times  U(1)_R$   doublet.  To   isolate  the  $U_A(1)$  breaking
contribution,  one can parameterize  the  two coupling constants $g_{\rm
sd}$ and  $g_{\rm   pd}$ as $g_{\rm  sd}  =  \lambda +  \delta \lambda$,
$g_{\rm pd} = -\lambda + \delta \lambda$.  Now, in the ladder truncation
scheme, it is    only $\delta \lambda$  that  can  provide the  $U_A(1)$
breaking effects to  \threeqBS{} amplitudes,  because $\lambda$ and  the
other $qq$  interactions  respect  the  $U_A(1)$ symmetry.    The   $qq$
interaction in the pseudo-scalar diquark channel  is often considered to
be irrelevant from    the non-relativistic  analogy.   However,  setting
$g_{\rm  pd}   =  0$  corresponds to  a  particular   choice of $U_A(1)$
breaking,  i.e., $\delta  \lambda =  g_{\rm  sd}/2$, and this particular
choice  is not   based  on any of    the underlying physics of  $U_A(1)$
breaking.   Due to this reason, we  suggest that the $qq$ interaction in
the  pseudoscalar diquark  channel  should be included  for a reasonable
estimate of the iso-scalar $g_A^{(0)}$, even if it is expected to give a
negligible contribution to the nucleon mass.
 
\begin{center}{\bf Acknowledgment}\end{center}
The author thanks K.  Yazaki, W.  Bentz, H.  Asami, L.  v. Smekal and H.
Terazawa for their  extensive  discussions and encouraging  suggestions.
He also thanks the  unknown referee  of his previous  paper\cite{ishii4}
for giving him the main motivation for the current work.
\appendix
\newpage
\section*{Appendices}
\section{The Faddeev Equation}
\label{appendix.faddeev.equation}

The    aim of  this  appendix  is  to  summarize  the  notations  of the
relativistic Faddeev equation and to provide the  derivations of some of
the relevant relations involving  the Faddeev method which are essential
to the other  parts of this paper  in a self-contained  manner.  In this
appendix, {\em summations over repeated  indices are not implied},  and,
unless explicitly indicated, {\em the external  fields are understood to
be absent}, i.e., $v\equiv a\equiv 0$, $m \equiv m_0$.

\subsection{The Relativistic Faddeev Equation\protect\footnote{A rather
good  pedagogical   introduction to the  Faddeev   equation is  found in
\protect\cite{afnan-tomas}.} and the Green's Function}
We    begin         with      the    \threeqBS{}       equation     (see
\Eq{3q.bs.equation.in.the.presence.of.external.field}) in the absence of
the external fields:
\begin{equation}
    G = G_0 + KG;
\Hs
    K = K_1 + K_2 + K_3,
\label{3q.bs.equation.in.the.absence.of.external.field}
\end{equation}
where the index $i$ of $K_i$ refers  to the spectator quark.  The formal
solution of $G$ is expressed by using the resolvent of $K$ as follows:
\begin{equation}
    G = {1\over 1 - K} G_0.
\label{formal.solution}
\end{equation}
Note that the  resolvent  exists  in  a mathematical  sense.    However,
because $K$ is an unbounded operator, it is difficult to interpret it as
it stands. Therefore, we  adopt the Faddeev prescription.   We introduce
the following Faddeev decomposition of the Green's function:
\begin{equation}
    G = G_0 + G^1 + G^2 + G^3;
\Hs
    G^i \equiv K_i G.
\label{def.gi}
\end{equation}
We    insert the    following     resolvent   identity  of  $K$     into
\Eq{formal.solution}:
\begin{equation}
    {1\over 1 -  K}
=
    {1  \over 1 - K_i}
+
    {1\over 1 - K_i}\left(\Tate \sum_{j\neq i} K_j \right)  {1\over 1 - K}.
\end{equation}
We obtain
\begin{eqnarray}
    G = {1\over 1 - K_i} G_0 + {1\over 1 - K_i} \sum_{j\neq i} G^j,
\end{eqnarray}
which   is further  inserted  into the  defining  relation   of $G^i$ in
\Eq{def.gi}.   We are   left  with the   following  closed equations for
$G^1,G^2,G^3$ (the Faddeev equations):
\begin{equation}
    G^i
=
    {K_i\over 1 - K_i} G_0
+
    {K_i\over 1 - K_i} \sum_{j\neq i} G^j.
\label{faddeev.eq}
\end{equation}
It is possible to simplify these coupled equations  into a single closed
equation for $G^3$ by using  the identical particle  nature of the three
quarks. $G^1$  and $G^2$  are obtained  from  $G^3$ by  means  of simple
permutation operations.  For this purpose, it is convenient to introduce
the cyclic permutation operator $Z$, which is defined as follows:
\begin{equation}
    (Z\psi)(x_1,x_2,x_3) = \psi(x_2,x_3,x_1).
\end{equation}
Note that $Z$ is easily implemented by using delta functions.  We give a
list of obvious relations:
\begin{equation}
\begin{array}[t]{l}
    Z^3 = 1,
\Hs
    Z(1 + Z + Z^2) = (1 + Z + Z^2)Z = 1 + Z + Z^2,
\\\displaystyle\Tate
    K_i = Z^i K_3 Z^{-i},
\Hs
    Z G = G Z = G,
\Hs
    Z G_0 = G_0 Z = G_0,
\Hs
    G^i = Z^i G^3.
\end{array}
\end{equation}
Now \Eq{faddeev.eq} reduces to   the following closed integral  equation
for $G^3$ (the reduced Faddeev equation):
\begin{equation}
    G^3
=
    {K_3 \over 1 - K_3} G_0
+
    {K_3 \over 1 - K_3} (Z + Z^2) G^3,
\label{faddeev.eq.for.g3}
\end{equation}
which is  depicted  in  \Fig{fig.faddeev}.(a).  The two-quark  resolvent
$\displaystyle     {K_3   \over   1    -     K_3}$  is    depicted    in
\Fig{fig.faddeev}.(b). (cf. \Eq{k=g0v}) It  is not  so hard to  identify
the  so-called  ``Z-diagram''  structure (the quark  exchange  diagram),
which is provided by  a combination of  the permutation operator $Z$ (or
$Z^2$) and   two external quark  propagators  of the two-quark resolvent
$\displaystyle {K_3\over 1  -  K_3}$.  We refer  to the  kernel of  this
integral equation as the Faddeev kernel $K_F$:
\begin{equation}
\label{eq.79}
    K_F \equiv {K_3\over  1 -  K_3} (Z + Z^2).
\end{equation}
Now the formal solution of $G^3$ to \Eq{faddeev.eq.for.g3} is given as:
\begin{equation}
    G^3
=
    {1\over 1 - K_F}
    {K_3 \over 1 - K_3}
    G_0.
\end{equation}
By inserting  this  into the   first  relation in \Eq{def.gi},  we  have
another formal representation of the Green's function $G$:
\begin{equation}
    G
=
    G_0
+
    (1 + Z + Z^2)
    {1\over 1 - K_F}
    {K_3 \over 1 - K_3}
    G_0.
\label{faddeev.representation.of.g}
\end{equation}

\subsection{The \threeqBS{}  Amplitude and the Faddeev Amplitude}

To obtain the form of the Green's function $G$ near the nucleon pole, we
first diagonalize  the  Faddeev kernel  $K_F$, regarding the  total four
momentum $p_{\mu}$  as  a parameter\footnote{The numerical  procedure to
solve  this  homogeneous Faddeev equation   is explained into  detail in
\cite{ishii1}.}:
\begin{equation}
    K_F \psi[n;p] = \lambda_n(p^2) \psi[n;p],
\Hs
    \Bar\psi[n;p] K_F = \lambda_n(p^2) \Bar \psi[n;p].
\end{equation}
We adopt the following normalization condition:
\begin{equation}
    \Bar \psi[n';p'] \psi[n;p]
=
    {\cal N}_n(p^2) (2\pi)^4 \delta^{(4)}(p'-p) \delta_{n'n};
\Hs
    {\cal N}_n(p^2) \equiv {i\over 2\sqrt{p^2} \lambda'_n(p^2)},
\label{normalization.condition}
\end{equation}
where $\lambda'_n(p^2) \equiv  d\lambda_n(p^2)/d(p^2)$. Now  the Faddeev
kernel is expressed as:
\begin{equation}
    K_F
=
    \sum_n \int {d^4 p\over (2\pi)^4 {\cal N}_n(p^2)}
    \lambda_n(p^2) \psi[n;p]\Bar\psi[n;p].
\end{equation}
Note  that, in  reasonable cases,  the  eigen-modes associated  with the
largest  eigenvalue are four-fold   degenerate ---they correspond to the
nucleon (the ground states  in the sector of  the baryon number 1).  The
degeneracy is due to  the iso-spin  1/2  and spin up/down.  We  refer to
these eigen-modes as  the (off-shell) eigenvalues and Faddeev amplitudes
for   the     nucleon   and  denote    them   as     $\lambda_{N}(p^2)$,
$\psi[N^a_{\alpha}(p)]$ and $\Bar\psi[N^a_{\alpha}(p)]$.    The  nucleon
mass $m_N$ is obtained by solving the following equation:
\begin{equation}
    \lambda_N(p^2 = m_N^2) = 1.
\end{equation}
The associated eigenvectors satisfy the homogeneous Faddeev equations:
\begin{equation}
    \psi[N^c_\beta(\vec L)]
=
    {K_3 \over 1 - K_3} (Z + Z^2)
    \psi[N^c_\beta(\vec L)],
\Hs
    \Bar\psi[N^a_{\alpha}(\vec P)]
=
    \Bar\psi[N^a_\alpha(\vec P)]
    {K_3 \over 1 - K_3} (Z + Z^2).
\label{homogeneous.faddeev.eq}
\end{equation}
For simplicity,  we  suppress the  time  components  of the  total  four
momenta  in the on-shell Faddeev  amplitudes.  We define the \threeqBS{}
amplitudes for nucleon states by:
\begin{eqnarray}
    \Psi[N^a_{\alpha}(\vec P)]
&\equiv&
    (1 + Z + Z^2)
    \psi[N^a_{\alpha}(\vec P)]
\label{def.3q.bs.amplitude}
\\\nonumber
    \Bar\Psi[N^a_{\alpha}(\vec P)]
&\equiv&
    \Bar\psi[N^a_{\alpha}(\vec P)]
    {K_3 \over 1 - K_3}
    G_0
\\\nonumber
&=&
    \Bar\psi[N^a_{\alpha}(\vec P)]
    {K_3 \over 1 - K_3}
    {1 + Z + Z^2 \over 3}
    G_0.
\end{eqnarray}
In order to see that these definitions of the \threeqBS{} amplitudes are
reasonable   and   that   the    normalization    scheme adopted      in
\Eq{normalization.condition}    is    consistent   with the    covariant
normalization of  the ket  vectors in  \Eq{covariant.normalization},  we
first consider the form of the Green's function near the nucleon pole as
follows (cf. \Eq{faddeev.representation.of.g}):
\begin{eqnarray}
    G
&\simeq&
    (1 + Z + Z^2)
    \sum_{a,\alpha}
    \int {d^4 P\over (2\pi)^4 {\cal N}_N(P^2)}
    {
	\psi[N^a_{\alpha}(P)] \Bar\psi[N^a_{\alpha}(P)]
    \over
	1 - \lambda_N(P^2)
    }
    {K_3 \over 1 - K_3}
    G_0
\\\nonumber
&\simeq&
    \sum_{a,\alpha}
    \int {d^4 P\over (2\pi)^4 {\cal N}_N(P^2)}
    \Psi[N^a_{\alpha}(\vec P)]
    { 1 \over 1 - \lambda_N(P^2) }
    \Bar\Psi[N^a_{\alpha}(\vec P)]
\\\nonumber    
&\simeq&
    \sum_{a,\alpha}
    \int {d^4 P\over (2\pi)^4 {\cal N}_N(m_N^2)}
    \times
    {1\over - \lambda'_N(m_N^2)}
    {
	\Psi[N^a_{\alpha}(\vec P)] \Bar\Psi[N^a_{\alpha}(\vec P)]
    \over
	P^2 - m^2 + i\epsilon
    }
\\\nonumber
&\simeq&
    \sum_{a,\alpha}
    \int {d^4 P\over (2\pi)^4}
    {m_N \over E_N(\vec P^2)}
    {i \over P_0 - E_N(\vec P^2) + i\epsilon}
    \Psi[N^a_{\alpha}(\vec P)] \Bar\Psi[N^a_{\alpha}(\vec P)].
\end{eqnarray}
This result is consistent with the expression which is expected from the
canonical  operator  analysis.  Note that  the   above manipulations are
exact at the nucleon pole.

Next, we show that these  \threeqBS{} amplitudes satisfy the homogeneous
\threeqBS{}            equation              \Eq{homogeneous.\threeqBS{}
.equation.without.external.field}:
\begin{eqnarray}
    K\Psi[N^a_\alpha(\vec P)]
&=&
    \sum_{i=1,2,3} Z^i K_3 Z^{-i}
    (1 + Z + Z^2) \psi[N^a_\alpha(\vec P)]
\label{eq.86}
\\\nonumber
&=&
    (1 + Z + Z^2) K_3
    (1 + Z + Z^2) \psi[N^a_\alpha(\vec P)]
\\\nonumber
&=&
    (1 + Z + Z^2) {K_3\over 1 - K_3} (Z + Z^2) \psi[N^a_\alpha(\vec P)]
=
    \Psi[N^a_\alpha(\vec P)]
\\
    \tildebar{\Psi}[N^a_\alpha(\vec P)] K
&=&
    \Bar\psi[N^a_\alpha(\vec P)]
    {K_3 \over 1 - K_3}
    {1 + Z + Z^2 \over 3}
    \sum_{i=1,2,3} Z^i K_3 Z^{-i}
\label{eq.87}
\\\nonumber
&=&
    \Bar\psi[N^a_\alpha(\vec P)]
    {K_3 \over 1 - K_3} (1 + Z + Z^2)
    K_3 {1 + Z + Z^2 \over 3}
\\\nonumber
&=&
    \Bar\psi[N^a_\alpha(\vec P)]
    {K_3 \over 1 - K_3}
    { 1 + Z + Z^2 \over 3}
=
    \tildebar\Psi[N^a_\alpha(\vec P)],
\end{eqnarray}
where, to obtain the 3rd lines of \Eq{eq.86} and \Eq{eq.87}, we used the
following identities:
\begin{equation}
\begin{array}[t]{l} \displaystyle
    \Bar \psi {K_3 \over 1 - K_3} (1 + Z + Z^2)
=
    \Bar \psi {K_3 \over 1 - K_3} + \Bar\psi
=
    \Bar \psi {1\over 1 - K_3}
\\\displaystyle \Tate
    (1 + Z + Z^2) \psi
=
    {K_3 \over 1 - K_3} (Z + Z^2) \psi
+   (Z + Z^2) \psi
=
    {1\over 1 - K_3} (Z + Z^2) \psi,
\end{array}
\label{faddeev.identity}
\end{equation}
which  immediately  follow   from  the   homogeneous   Faddeev equations
\Eq{homogeneous.faddeev.eq}.

\subsection{The Matrix Elements in terms of the Faddeev Amplitude}
\label{appendix.a.3}

Here, we  derive an explicit  expression for the  matrix  element of the
axial vector current operator in  terms of the Faddeev amplitudes, which
is, in the practical applications,  more convenient than the  expression
in terms of the  \threeqBS{} amplitudes.  To avoid cumbersome notations,
we  introduce a  shorthand  notation:  $\displaystyle O_{i;\mu}^b \equiv
\left[{i\delta    K_i^{[e]}      \over \delta a_{\mu}^b(x     =  0)}
\right]^{[0]}$ (i=1,2,3) and  $\displaystyle O_{\mu}^b =  \sum_{i=1,2,3}
O_{i;\mu}^b$, where the index  $i$ refers to  the spectator quark.  Note
that $O_{i;\mu}^b  =    Z^i O_{3;\mu}^b  Z^{-i}$.   Now  we  have   from
\Eq{matrix.element.1}:
\begin{eqnarray}
\lefteqn{
    \left\langle N^a_{\alpha}(\vec P) \left| \Tate
	A_{\mu}^b(x=0)
    \right| N^c_{\beta}(\vec L) \right\rangle
}
\nonumber
\\\nonumber
&=&
    \tildebar{\Psi}[N^a_{\alpha}(\vec P)]
    O^b_{\mu}
    \Psi[N^c_{\beta}(\vec L)]
\\\nonumber
&=&
    \Bar\psi[N^a_{\alpha}(\vec P)]
    {K_3 \over 1 - K_3}
    (1 + Z + Z^2)
    O^b_{3;\mu}
    (1 + Z + Z^2)
    \psi[N^c_{\beta}(\vec L)]
\\\nonumber
&=&
    \Bar\psi[N^a_\alpha(\vec P)]
    {1 \over 1 - K_3} O^b_{3;\mu} {1\over 1 - K_3} (Z + Z^2)
    \psi[N^c_\beta(\vec L)]
\\
\label{matrix.element.faddeev.equation}
&=&
    \Bar\psi[N^a_\alpha(\vec P)]
    \left[{i \delta K_F^{[e]} \over \delta a^a_{\mu}(x=0)}\right]^{[0]}
    \psi[N^c_\beta(\vec L)],
\end{eqnarray}
where, to obtain the fourth line, we  used \Eq{faddeev.identity}.  Since
the permutation operator  $Z$ does  not depend  on the  external fields,
$\delta/\delta  a_{\mu}^b$    only   hits  one   the   constituent quark
propagators in   the two-quark resolvent $\displaystyle  {K_3  \over 1 -
K_3}$ in $K_F$. (cf. \Eq{eq.79})

We comment   on the ladder  truncation scheme.   In this  case, the $qq$
interaction is point-like.   By combining the diagramatic expressions of
$K_3$ [\Fig{fig3}.(a)] and  the two-quark resolvent $\displaystyle  {K_3
\over 1    - K_3}$ [\Fig{fig.faddeev}.(b)],   we see  that  the diagrams
involved in $i   \delta  K_F/\delta a^a_{\mu}$ are  classified  into two
types,  i.e., (1)  $\delta/\delta a^a_{\mu}$  hits  one of  the internal
quark propagators (i.e., in the $qq$ bubble diagram)  in the ladder sum,
(2)   $\delta/\delta a^a_{\mu}$ hits  one    of the two external   quark
propagators.  With the aid   of this classification, a  straight forward
diagramatic  argument   shows   that  the    diagramatic expression   of
\Eq{matrix.element.faddeev.equation}  is   given by  the  three diagrams
\Fig{fig7}.(a), \Fig{fig7}.(b) and \Fig{fig7}.(c).  The first type leads
to   \Fig{fig7}.(c),  which we  refer   to  as  the ``diquark  current''
contribution,  and   the   second  type  leads    to \Fig{fig7}.(a)  and
\Fig{fig7}.(b), which we refer to  as the ``quark current'' contribution
and the ``exchange current'' contribution, respectively.

\section{Notations of the NJL Model}
\label{convensions.of.the.njl.model}

The aims of this section are  to define some  of the relevant quantities
with explicit examples, and to summarize the notations of the NJL model.
In this section, {\em the external fields are  understood to be absent},
i.e., $v^a_{\mu}(x) \equiv a^a_{\mu}(x) \equiv 0$, $m(x) \equiv m_0$.

\subsection{The Elementary Effective Coupling Constants and the Closed
Chiral Multiplets}

Here we define the elementary  effective coupling constants in $q\bar q$
and     $qq$   channels.   We   start    with    the Lagrangian  density
\Eq{lagrangian}.  For definiteness,   we virtually distinguish  the  two
$\psi$'s and the two $\bar\psi$'s from each other, respectively, i.e.,
\begin{equation}
    {\cal L}_I
=
    \sum_{\Gamma}
    g_{\Gamma}
    (\bar\psi_1  \Gamma   \psi_2)
    (\bar\psi_3  \Gamma   \psi_4).
\end{equation}
The elementary interaction,  which  is depicted in  \Fig{fig.fierz}.(a),
can be  classified into the following three  types: (i) $q\bar q$ direct
channel  [\Fig{fig.fierz}.(b)],   (ii)    $q\bar   q$  exchange  channel
[\Fig{fig.fierz}.(c)], (iii) $qq$ diquark channel [\Fig{fig.fierz}.(d)].
They  are related by  Fierz  identities to  each  other.  We define  the
effective coupling constants   $g^{(q\bar q {\rm  dir})}$, $g^{(q\bar  q
{\rm exch})}$ and $g^{(qq)}$ through the following relations:
\begin{eqnarray}
    {\cal L}_I
&=&
    \sum_{\alpha}
    \sum_{i=0}^3
    \sum_{A=0}^8
    g^{(q\bar q {\rm dir})}_{\alpha i A}
    \left(\Tate \bar\psi_1 \Gamma_\alpha \tau_i \beta_A \psi_2 \right)
    \left(\Tate \bar\psi_3 \Gamma^\alpha \tau_i \beta_A \psi_4 \right)
\\\nonumber
&=&
    \sum_{\alpha}
    \sum_{i=0}^3
    \sum_{A=0}^8
    g^{(q\bar q {\rm exch})}_{\alpha i A}
    \left(\Tate \bar\psi_1 \Gamma_\alpha \tau_i \beta_A \psi_4 \right)
    \left(\Tate \bar\psi_3 \Gamma^\alpha \tau_i \beta_A \psi_2 \right)
\\\nonumber
&=&
    \sum_{\alpha}
    \sum_{i=0}^3
    \sum_{A=0}^8
    g^{(qq)}_{\alpha i A}
    \left(\Tate
	\bar\psi_1
	(\Gamma_\alpha \gamma_5 C^{-1})
	(\tau_i\tau_2)
	\beta_A
	\bar\psi_3^T
    \right)
    \left(\Tate
	\psi_2^T
	(C\gamma_5\Gamma^\alpha)
	(\tau_2\tau_i)
	\beta_A
	\psi_4
    \right).
\end{eqnarray}
$\Gamma_\alpha,\Gamma^{\alpha}$ are Dirac gamma matrices, where $\alpha$
runs   over    S(scalar),  V(vector),   T(tensor),  A(axial-vector)  and
P(pseudo-scalar), i.e.,
\begin{eqnarray}
    (\Gamma_{\alpha})_{\alpha=\rm S,V,T,A,P}
&=&
    (1,\gamma_{\mu},\sigma_{\mu\nu},\gamma_{\mu}\gamma_5,\gamma_5)
\\\nonumber
    (\Gamma^{\alpha})_{\alpha=\rm S,V,T,A,P}
&=&
    (1,\gamma^{\mu},\sigma^{\mu\nu},\gamma_5\gamma^{\mu},\gamma_5).
\end{eqnarray}
$\tau_i$ ($i=0,1,2,3$) are  the iso-spin Pauli  matrices  with $\tau_0 =
1$.    They  are normalized    according  to $\mbox{tr}(\tau_i\tau_j)  =
2\delta_{ij}$.   Note that, both in  $q\bar q$ and $qq$ representations,
$\tau_0$ corresponds  to   the iso-scalar,   and  $\tau_i$   ($i=1,2,3$)
correspond to  the iso-vector   channels.  $\beta_A$  are   the rescaled
Gell-Mann color matrices, i.e., $\beta_0 =  1$, $\displaystyle \beta_A =
\sqrt{3\over 2} \lambda_A$  ($A=1,2,\cdots,  8$) with  the normalization
$\mbox{tr}(\beta_A\beta_B)    =   3    \delta_{AB}$.    In $q\bar     q$
representations, $\beta_0$  corresponds to the  $1_c$ mesonic  channels,
and   $\beta_A$  for $A=1,2,\cdots,8$     corresponds to  $8_c$  mesonic
channels.  Note that $\beta_A$ for  $A=2,5,7$ are anti-symmetric and the
others are symmetric.   Therefore, in $qq$ representation, $\beta_A$ for
$A=2,5,7$ correspond to $\bar  3_c$ diquark channels, and $\beta_A$  for
$A=0,1,3,4,6,8$ correspond   to $6_c$ diquark  channels.  We  define the
effective coupling constants $g^{(q\bar q)}$ as follows:
\begin{equation}
    g^{(q\bar q)}_{\alpha i A}
=
    g^{(q\bar q {\rm dir})}_{\alpha i A}
+   g^{(q\bar q {\rm exch})}_{\alpha i A}.
\end{equation}
We consider  two   examples   (i) the  original NJL    type  interaction
Lagrangian:
\begin{equation}
    {\cal L}_I
=
    g\left(\Tate
	(\bar\psi\psi)^2 - \sum_{i=1}^3(\bar\psi\gamma_5\tau_i\psi)^2
    \right),
\end{equation}
where $\tau_i$  is the isospin Pauli  matrix, and (ii) the color current
type interaction Lagrangian:
\begin{equation}
    {\cal L}_I
=
    -g\sum_{a=1}^8
    \left(\Tate \bar\psi \gamma_{\mu} {\lambda_a\over 2}\psi \right)^2,
\end{equation}
where  $\lambda_a$ is the  color Gell-Mann  matrix.  We  give a  list of
these   effective coupling  constants $g^{(q\bar q)}   = g^{(q\bar q{\rm
dir})}  +   g^{(q\bar q{\rm  exch})}$ and   $g^{(qq)}$   for  these  two
interaction Lagrangians:
\begin{center}
\begin{tabular}{|l|r|r|r|r|r|}
\hline\hline
\Tate {\bf Original NJL Type} & {\bf S} & {\bf V} & {\bf T} & {\bf A} & {\bf P} \\
\hline\hline
$1_c$ ($q\bar q$), $I=0$
    & $g + g/12$ & $0 -g/6$ & $0 + g/12$ & $0 - g/6$ & $0 + g/12$ \\
\hline
$1_c$ ($q\bar q$), $I=1$
    & $0 -g/12$& $0 + 0$& $0-g/12$& $0 + 0$& $-g -g/12$\\
\hline
$8_c$ ($q\bar q$), $I=0$
    & $0 + g/12$& $0-g/6$& $0+g/12$& $0-g/6$& $0+g/12$ \\
\hline
$8_c$ ($q\bar q$), $I=1$
    & $0-g/12$& $0+0$& $0-g/12$& $0+0$& $0-g/12$\\
\hline\hline
\Tate $\bar 3_c$ ($qq$), $I=0$ & $g/6$ & $-g/12$& $g/6$ \tozero & $-g/12$ \tozero & $g/6$\\
\hline
\Tate $\bar 3_c$ ($qq$), $I=1$ & 0 \tozero & $g/12$ \tozero & 0& $g/12$& 0 \tozero\\
\hline
$6_c$ ($qq$), $I=0$ & $g/6$ \tozero & $-g/12$ \tozero & $g/6$ & $-g/12$ & $g/6$ \tozero\\
\hline
$6_c$ ($qq$), $I=1$ & 0 & $g/12$& 0 \tozero & $g/12$ \tozero & 0\\
\hline\hline\hline
\Tate {\bf Color Current Type} & {\bf S} & {\bf V} & {\bf T} & {\bf A} & {\bf P} \\
\hline\hline
$1_c$ ($q\bar q$), $I=0$ & $0+2g/9$ & $0-g/9$ & $0+0$ &$0+g/9$ &$0-2g/9$ \\
\hline
$1_c$ ($q\bar q$), $I=1$ & $0+2g/9$& $0-g/9$& $0+0$& $0+g/9$ & $0-2g/9$\\
\hline
$8_c$ ($q\bar q$), $I=0$ & $0-g/36$& $-g+g/72$ & $0+0$ & $0-g/72$ & $0+g/36$\\
\hline
$8_c$ ($q\bar q$), $I=1$ & $0-g/36$& $0+g/72$ & $0+0$ & $0-g/72$ & $0+g/36$\\
\hline\hline
\Tate $\bar 3_c$ ($qq$), $I=0$ & $g/9$& $-g/18$ & 0 \tozero & $g/18$ \tozero & $-g/9$ \\
\hline
\Tate $\bar 3_c$ ($qq$), $I=1$ & $g/9$ \tozero& $-g/18$ \tozero & 0 & $g/18$ & $-g/9$ \tozero \\
\hline
$6_c$ ($qq$), $I=0$ & $-g/18$ \tozero & $g/36$ \tozero & 0 & $-g/36$ & $g/18$ \tozero\\
\hline
$6_c$ ($qq$), $I=1$ & $-g/18$& $g/36$ & 0 \tozero & $-g/36$ \tozero & $g/18$\\
\hline\hline
\end{tabular}
\end{center}
For completeness, we also presented the effective couplings in the $6_c$
diquark channels, which do not contribute directly  to the color singlet
baryon states.  Note that  the two $\bar\psi$  fields and the two $\psi$
fields are  originally undistinguished Grassmann  fields.  Therefore, in
$qq$ representation, unless $\Gamma_\alpha \gamma_5 C^{-1} \tau_i \tau_2
\beta_A$ is anti-symmetric,  the  contribution  vanishes.  Due  to  this
reason, half of  the effective coupling  constants in  the $qq$ channels
actually vanish, and  these cases are  indicated  by ``\tozero'' in  the
list.

Due  to the chiral $SU(2)_L\times SU(2)_R$   symmetry of the interaction
Lagrangian, it can be already seen from the above list  that some of the
effective   coupling  constants  are grouped  together.     In fact, the
straight forward   application of  the  chiral  $SU(2)_L\times  SU(2)_R$
transformation leads  us  to   the  following list of    ``closed chiral
multiplets''\footnote{For the precise meaning, see Section.4.}:
\[
\begin{array}{||l|l||l|l||}
\hline\hline
\Tatetate
q\bar q (1_c) &	\left(\Tate V,I=0 \right)
& q\bar q (8_c) & \left(\Tate V,I=0 \right)
\\\Tatetate
&    \left(\Tate A,I=0 \right)		&&   \left(\Tate A,I=0 \right)       
\\\Tatetate
&    \left(\Tate (S,I=0)-(P,I=1)\right)	&&   \left(\Tate (S,I=0)-(P,I=1)\right)
\\\Tatetate
&    \left(\Tate (S,I=1)-(P,I=0)\right)	&&   \left(\Tate (S,I=1)-(P,I=0)\right)
\\\Tatetate
&    \left(\Tate (T,I=0)-(T,I=1)\right)	&&   \left(\Tate (T,I=0)-(T,I=1)\right)
\\\Tatetate
&    \left(\Tate (V,I=1)-(A,I=1)\right)	&&   \left(\Tate (V,I=1)-(A,I=1)\right)
\\\hline\hline
\Tatetate
qq (\bar 3_c) &	\left(\Tate S,I=0\right)
& qq (6_c) &     \left(\Tate S,I=1\right)
\\\Tatetate
&    \left(\Tate T,I=1\right)		&&   \left(\Tate T,I=0\right)          
\\\Tatetate
&    \left(\Tate P,I=0\right)		&&   \left(\Tate P,I=1\right)          
\\\Tatetate
&    \left(\Tate (V,I=0)-(A,I=1)\right)	&&   \left(\Tate (V,I=1)-(A,I=0)\right)
\\\hline\hline
\end{array}
\]
Note  that  the total number  of independent  coupling  constants of the
chiral $SU(2)_L\times SU(2)_R$ symmetric NJL type interaction Lagrangian
is at most eight.  Therefore, if we fix all the eight coupling constants
in the $qq$ channels, all the effective coupling constants in the $q\bar
q$  channels are obtained  as their linear  combinations.  In the ladder
truncation  scheme of  the \threeqBS{}  kernel, we can  fix the coupling
constants  in the $\bar 3_c$ ($qq$)  channels  based on the calculations
for color singlet baryons.  However, the coupling constants in the $6_c$
($qq$) channels remain free.  One can  then use these remaining coupling
constants (related to the $q\bar q$ coupling constants) to reproduce the
mesonic properties.

To avoid cumbersome notations, we introduce the following abbreviations:
\begin{equation}
    g_{\pi} = - g^{(q\bar q)}_{P,I=1,1_c}= g^{(q\bar q)}_{S,I=0,1_c},
\Hs
    g_{\rm sd} = g^{(qq)}_{S,I=0,\bar 3_c},
\Hs
    g_{\rm ax} = g^{(q\bar q)}_{A,I=1,1_c} = - g^{(q\bar q)}_{V,I=1,1_c}.
\end{equation}

\subsection{The Vacuum of the NJL model at the Mean Field Level}
Here, we summarize some of the notations of the NJL  model in the vacuum
and the mesonic sectors.

\subsubsection{$g_{\rm ax} = 0$ Case}

The gap equation is given by:
\begin{equation}
	M
=
	m_0
+
	2ig_{\pi} \int {d^4p\over (2\pi)^4}
	\Tr S_F(p),
\Hs
	S_F(p) \equiv {1\over \fslash{p} - M},
\label{explicit.gap.equation}
\end{equation}
which  provides the spontaneous breaking  of the chiral  symmetry in the
NJL  model  at the  level of   mean   field approximation.  $M$   is the
constituent  quark mass.  The pion mass  and the pion decay constant are
obtained from the two-point axial current correlator as follows:
\begin{eqnarray}
    \int d^4 x e^{iqx}
    \left\langle 0 \left|\Tate
	T A_{\mu}^a(x) A_{\nu}^b(0)
    \right| 0 \right\rangle
&=&
    {iq_{\mu}q_{\nu} f_{\pi}^2 \over q^2 - m_{\pi}^2 + i\epsilon}  
+
    \mbox{continuum}.
\end{eqnarray}
Here we adopted the following definition of the pion decay constant:
\begin{equation}
    \left\langle 0 \left|\Tate
	A_{\mu}^a(x)
    \right| \pi^b(\vec{p}) \right\rangle
=
    i p_{\mu} f_{\pi} \delta_{ab} e^{-ipx},
\end{equation}
where    the   covariant   normalization  $\left\langle   \pi^a(\vec  p)
\left|\Tate  \right.  \pi^b(\vec k) \right\rangle  = 2 \sqrt{m_{\pi}^2 +
\vec p^2}  (2\pi)^3 \delta^{(3)}(\vec    p -  \vec k)   \delta_{ab}$  is
adopted. The explicit form of the axial current correlator can be easily
obtained in the ladder approximation. However, in order to emphasize the
consistency between the   vacuum, the mesonic  sector and   the baryonic
sector in our  formulation, we prefer to  use the external field method.
Since the canonical operator expression of $\delta S_F/\delta a_{\nu}^b$
is
\begin{equation}
    \left[
        {i \delta i S_F^{[e]}(x,z)_{\alpha\beta} \over \delta a_{\nu}^b(y)}
    \right]^{[0]}
=
    \left\langle 0 \left| \Tate
	T \psi_{\alpha}(x) \Bar\psi_{\beta}(z) A_{\nu}^b(y)
    \right| 0 \right\rangle,
\end{equation}
the explicit form  of  the axial current correlator  is  obtained in the
following way(cf. eqs. (\ref{closed.eq.for.dsigma.da}) - (\ref{eq.37})):
\begin{eqnarray}
\lefteqn{
    \int d^4 x e^{iq(x - y)}
    \left\langle 0 \left| \Tate
	T A_{\mu}^a(x) A_{\nu}^b(y)
    \right| 0 \right\rangle
}
\\\nonumber
&=&
    \int d^4 x
    e^{iq(x - y)}
    \Tr\left(
	(\gamma_{\mu}\gamma_5 {\tau^a\over 2})
	\left[
	    {\delta S_F^{[e]}(x,x) \over \delta a^{b}_{\mu}(y)}
	\right]^{[0]}
    \right)
\\\nonumber
&=&
    -{1\over 4} \delta_{ab}\Pi_{\mu\nu}(q)
    -{1\over 4} q_{\mu}q_{\nu}\delta_{ab} \Pi_{5A}(q^2) \tilde H(q^2)
\\\nonumber
&=&
-
    {1\over 4}
    \Pi_{\mu\nu}(q)
    \delta_{ab}
-
    {1\over 4}
    q_{\mu}q_{\nu}\delta_{ab}
    {
	\Pi_{5A}(q^2) (-2ig_{\pi}) \Pi_{5A}(q^2)
    \over
	1 + 2ig_{\pi} \Pi_{55}(q^2)
    },
\end{eqnarray}
where
\begin{eqnarray}
    \Pi_{55}(q^2)\delta_{ab}
&\equiv&
    - \int {d^4k\over(2\pi)^4}
    \Tr\left(\Tate
	(\gamma_5\tau^a) iS_F(k + q) (\gamma_5\tau^b) iS_F(k)
    \right)
\\\nonumber
    q_{\mu}\Pi_{5A}(q^2)\delta_{ab}
&\equiv&
    - \int {d^4k\over(2\pi)^4}
    \Tr\left(\Tate
	(\gamma_5\tau^a)
	iS_F(k + q)
	(\gamma_{\mu}\gamma_5 {\tau^b})
	iS_F(k)
    \right)
\\\nonumber
&=&
    - \int {d^4k\over(2\pi)^4}
    \Tr\left(\Tate
	(\gamma_5\gamma_{\mu} {\tau^b})
	iS_F(k + q)
	(\gamma_5\tau^a)
	iS_F(k)
    \right)
\\\nonumber
    \Pi_{\mu\nu}(q)\delta_{ab}
&\equiv&
    - \int {d^4k\over(2\pi)^4}
    \Tr\left(\Tate
	(\gamma_\mu\gamma_5\tau^a)
	iS_F(k + q)
	(\gamma_5\gamma_{\nu}\tau^b)
	iS_F(k)
    \right)
\\\nonumber
&\equiv&
    \delta_{ab}
    \left({q_{\mu}q_{\nu}\over q^2}\right)
    \Pi_1(q^2)
+
    \delta_{ab}
    \left( g_{\mu\nu} - {q_{\mu}q_{\nu}\over q^2}\right)
    \Pi_2(q^2).
\end{eqnarray}

The explicit expressions for $m_\pi$ and $f_{\pi}$ are obtained from:
\begin{equation}
    0
=
    1 + 2ig_{\pi} \Pi_{55}(m_{\pi}^2),
\Hs
    f_{\pi}
=
    {-i\Pi_{5A}(m_{\pi}^2) \over 2\sqrt{ -i \Pi'_{55}(m_{\pi}^2)}},
\label{vacuum.parameter.1}
\end{equation}
where $\displaystyle   \Pi'_{55}(q^2) \equiv {d \Pi_{55}(q^2) \over d(q^2)}$.

\subsubsection{$g_{\rm ax} \neq 0$ Case}

Even  if we switch  on $g_{\rm ax}  \neq 0$, the  gap  equation does not
change. However, the axial current  correlator changes in the  following
way:
\begin{eqnarray}
\lefteqn{
    \int d^4 x e^{iq(x - y)}
    \left\langle 0 \left| \Tate
	T A_{\mu}^a(x) A_{\nu}^b(y)
    \right| 0 \right\rangle
}
\\\nonumber
&=&
    \Bs
    \begin{array}[t]{l}	\displaystyle
        -{1\over 4} \delta_{ab}
	\Pi_{\mu\nu}(q)
    -
       {1\over 4} \delta_{ab}
	\left( g_{\mu\nu} - {q_{\nu}q_{\mu} \over q^2} \right)
        \Pi_2(q^2) G_2(q^2)
    \\\displaystyle\Tatetate
    -   {1\over 4}\delta_{ab}
	\left( {q_{\mu}q_{\nu}\over q^2} \right)
	\left(\Tate q^2 H(q^2) \Pi_{5A}(q^2) + G(q^2) \Pi_1(q^2) \right)
    \end{array}
\\\nonumber
&=&
    -{1\over 4} \delta_{ab}
    \Pi_{\mu\nu}(q)
-
   {1\over 4} \delta_{ab}
    \left( g_{\mu\nu} - {q_{\nu}q_{\mu} \over q^2} \right)
    {2ig_{\rm ax} \Pi_2(q^2) \over 1 - 2ig_{\rm ax} \Pi_2(q^2)}
-   {1\over 4}\delta_{ab}
    \left( {q_{\mu}q_{\nu}\over q^2} \right)
    { N(q^2) \over D(q^2) },
\end{eqnarray}
where   $H(q^2)$,  $G_1(q^2)$, $G_2(q^2)$ and   $D(q^2)$  are defined in
\Appendix{appendix.non-vanishing.gax.gtv.index}, and $N(q^2)$ is   given
as follows:
\begin{eqnarray}
    N(q^2)
&=&
    \Bs
    \begin{array}[t]{l}	\displaystyle
	-{1\over 4} q^2 (-2ig_{\pi})
	\left(\Tate \Pi_{5A}(q^2) \right)^2
	\left(\Tate 1 + 2ig_{\rm ax} \Pi_1(q^2) \right)
    \\\displaystyle\Tatetate
    -	{1\over 4}
	\left(\Tate 1 + 2ig_{\pi} \Pi_{55}(q^2) \right)
	\left(\Tate 2ig_{\rm ax} \Pi_1(q^2) \right)
    \end{array}
\end{eqnarray}
The explicit expressions for $m_{\pi}$ and $f_{\pi}$ are obtained from:
\begin{equation}
    0 = D(m_{\pi}^2),
\Hs
    f_{\pi} = \sqrt{ N(m_{\pi}^2) \over i m_{\pi}^2 D'(m_{\pi}^2) },
\label{vacuum.parameter.2}
\end{equation}
where $\displaystyle D'(q^2) \equiv {d D(q^2) \over d(q^2)}$.

\subsection{ The Regularized Bubble Integrals }

The regularized   expressions for the bubble   integrals  with the sharp
Euclidean cut-off are as follows:
\begin{eqnarray}
    \Pi_{55}(q^2)
&=&
    24 i
    \int^{\Lambda} {d^4 k_E\over(2\pi)^4}
    \int_0^1 dx
    {
	k^2_E + M^2 + q^2 x(1 - x)
    \over
	(k^2_E + M^2 - q^2 x(1 - x))^2
    }
\\
    \Pi_{5A}(q^2)
&=&
    24 i M
    \int^{\Lambda} {d^4 k_E\over(2\pi)^4}
    \int_0^1 dx
    {
	1
    \over
	(k^2_E + M^2 - q^2 x(1 - x))^2
    }
\\
    \Pi_{\mu\nu}(q)
&=&
    48 i
    \int^{\Lambda} {d^4 k_E\over(2\pi)^4}
    \int_0^1 dx
    {
	M^2 g_{\mu\nu}
    -	x(1 - x)\left(\Tate q^2 g_{\mu\nu} - q_{\mu}q_{\nu}\right)
    \over
	(k^2_E + M^2 - q^2 x(1 - x))^2
    }.
\end{eqnarray}
These expressions satisfy the following identities:
\begin{eqnarray}
    \Pi_{5A}(0)
&=&
    2M\Pi'_{55}(0)
\label{ward.identity.1}
\\
    q^{\nu}\Pi_{\mu\nu}(q)
&=&
    2Mq_{\mu}\Pi_{5A}(q^2)
\label{ward.identity.2}
\\
    \Pi_1(0)
&=&
    \Pi_2(0).
\label{ward.identity.3}
\end{eqnarray}
We  comment here on the  expression for  $\Pi_{\mu\nu}(q)$.  In order to
obtain the  above expression for $\Pi_{\mu\nu}(q)$, we  have to  use the
following prescription:
\begin{equation}
    \int {d^4k\over(2\pi)^4}
    \int_0^1 dx
    {
	2k_{\mu}k_{\nu}
    +	g_{\mu\nu}\left(\Tate M^2 - k^2 - q^2 x(1 - x)\right)
    \over
	(M^2 - k^2 - q^2 x(1 - x))^2
    }
\Rightarrow
    0
\label{gauge.invariant.prescription}
\end{equation}
This  is due to the following  reason.  The straight forward application
of     the  sharp  Euclidean      cut-off    leads,  rather   than    to
\Eq{ward.identity.2}, to the following identity:
\begin{equation}
    q^{\nu} \Pi_{\mu\nu}(q)
=
    -q^{\nu} \Pi^{(V)}_{\mu\nu}(q)
+   2Mq_{\mu} \Pi_{5A}(q^2),
\label{ward.identity.2a}
\end{equation}
where $\Pi^{(V)}_{\mu\nu}(q)$ is
\begin{equation}
    \Pi^{(V)}_{\mu\nu}(q)\delta_{ab}
\equiv
    -
    \int {d^4k\over(2\pi)^4}
    \Tr\left(\Tate
	(\gamma_{\mu}\tau^a)
	iS_F(k+q)
	(\gamma_{\nu}\tau^b)
	iS_F(k)
    \right).
\end{equation}
Since the  non-vanishing longitudinal part of $\Pi^{(V)}_{\mu\nu}(q)$ is
an  unphysical cutoff artifact, which  spoils   the significance of  the
outputs, the   prescription  \Eq{gauge.invariant.prescription} is  often
used to  suppress it.  (See p.178 in  ref.  \cite{huang}.) Note that, if
the    dimensional      regularization is   applied,     the  expression
\Eq{gauge.invariant.prescription}     vanishes  identically.   Now   the
analogous   subtraction  should   be performed    on   the   l.h.s.   in
\Eq{ward.identity.2a},    otherwise the Gell-Mann-Oakes-Renner  relation
would be terribly violated in the case $g_{\rm ax}  \neq 0$.  Note that,
once we use this prescription, the Gell-Mann-Oakes-Renner violations are
less than 1\% for $m_{\pi} \le 140$ MeV.

\section{Matrix Element in terms of the \threeqBS{}/Faddeev Amplitudes in
the Rest Frame}

In principle, we can use  any Lorentz frames   to calculate the  Lorentz
invariant form factors  of  a matrix  element.  But in  practice,  it is
convenient to  make use of the  rest frame, because  this frame is often
used in the calculation of the mass and the \threeqBS{} amplitudes.  The
aim  of this appendix is to  give expressions for  the matrix element in
terms of the \threeqBS{} /Faddeev amplitudes  in the rest frame by using
the Lorentz covariance.   Unless  the opposite is explicitly  indicated,
{\em the external  fields are understood  to be  absent}, i.e., $v=a=0$,
$m=m_0$.   To make the Lorentz transformation  of the spinor simpler, we
select  the final  momentum  $P=(m_N,\vec{0})$ and the initial  momentum
$L^{\mu}={(\Lambda^{-1})^{\mu}}_{\nu}             P^{\nu}$,        where
${\Lambda^{\mu}}_{\nu} = {(e^\omega)^{\mu}}_{\nu}$ represents the boost.
We introduce the Fourier transforms of the \threeqBS{} amplitudes and of
$O^b_\mu\equiv\left[  \displaystyle{i    \delta  K^{[e]}    \over \delta
a_{\mu}^b(0) } \right] \Zero $:
\begin{eqnarray}
\lefteqn{
	\Psi[N^c_\beta(L)](l,l')\;
	(2\pi)^4 \delta^{(4)}(L - \bar{L})
}
\label{fourier.transform}
\\
\nonumber
&=&
	\int d^4 y_1 d^4 y_2 d^4 y_3\;
	e^{-i\sum l_i y_i}
	\Psi[N^c_\beta(L)](y_1,y_2,y_3)
\\
\nonumber
\lefteqn{
	\tildebar{\Psi}[N^a_\alpha(P)](p,p')\;
	(2\pi)^4 \delta^{(4)}(\bar{P} - P)
}
\\
\nonumber
&=&
	\int d^4 x_1 d^4 x_2 d^4 x_3\;
	\tildebar{\Psi}[N^a_\alpha(P)](x_1,x_2,x_3)\;
	e^{i\sum p_i x_i}
\\
\nonumber
\lefteqn{
	O^b_{\mu}(\bar{P},p,p';\bar{L},l,l')
}
\\
\nonumber
&=&
	\int d^4 x_1 d^4 x_2 d^4 x_3
	\int d^4 y_1 d^4 y_2 d^4 y_3\;
	e^{-i\sum p_i x_i}\;
	O^b_\mu(x_1,x_2,x_3;y_1,y_2,y_3)\;
	e^{i\sum l_i y_i}
\\
\nonumber
\lefteqn{
	O^b_\mu(P,L;p,p',l,l')
}
\\
\nonumber
&=&
	\int {d^4 \bar{P}\over (2\pi)^4}
	\int {d^4 \bar{L}\over (2\pi)^4}\;
	(2\pi)^4 \delta^{(4)}(\bar{P} - P)\;
	O^b_{\mu}(\bar{P},p,p';\bar{L},l,l')
	(2\pi)^4 \delta^{(4)}(L - \bar{L})
\end{eqnarray}
$\bar{P}, \bar{L}$   are  the total momenta   and $p,p',l,l'$   are  the
relative momenta, which are defined though the following relations:
\begin{equation}
\begin{array}[t]{lll}
	\bar{P} = \sum p_i
&
	p = \eta p_3 - (1 - \eta)(p_1 + p_2),
&
	p' = \eta' p_1 - (1 - \eta')p_2
\\
	\Rule{1.2\Tatescale}
	\bar{L} = \sum l_i
&
	l = \eta l_3 - (1 - \eta)(l_1 + l_2),
&
	l' = \eta' l_1 - (1 - \eta')l_2,
\end{array}
\label{relative.total.momentum}
\end{equation}
where $0  < \eta,\eta' < 1$ are  arbitrary real  numbers.  We note that,
unlike the nonrelativistic approaches, the choice  of $\eta$ and $\eta'$
is  almost  completely  arbitrary  in  the   relativistic quantum  field
theory\footnote{For  complete   expositions, see \cite{itzykson}.}.  The
delta functions in the first two relations in \Eq{fourier.transform} are
due to the  translational invariance.  It is important  to note that the
Jacobians  associated with the   variable change,  i.e.,  $(p_1,p_2,p_3)
\mapsto (\bar{P},p,p')$ and  $(l_1,l_2,l_3) \mapsto  (\bar{L},l,l')$ are
1, i.e.,
\begin{equation}
	{d^4 p_1 \over (2\pi)^4}
	{d^4 p_2 \over (2\pi)^4}
	{d^4 p_3 \over (2\pi)^4}
=
	{d^4 \bar{P} \over (2\pi)^4}
	{d^4 p \over (2\pi)^4}
	{d^4 p' \over (2\pi)^4},
\Hs
	{d^4 l_1 \over (2\pi)^4}
	{d^4 l_2 \over (2\pi)^4}
	{d^4 l_3 \over (2\pi)^4}
=
	{d^4 \bar{L} \over (2\pi)^4}
	{d^4 l \over (2\pi)^4}
	{d^4 l' \over (2\pi)^4}.
\end{equation}
Now the matrix element reduces to the following expression:
\begin{eqnarray}
\lefteqn{
	\left\langle N^a_\alpha(P)\left| \Tate
		A_{\mu}^b(0)
	\right| N^c_\beta(L)\right\rangle
\left(
\equiv
	\tildebar{\Psi}[N^a_\alpha(P)]\;
	O^b_\mu\;
	\Psi[N^c_\beta(L)]
\right)
}
\label{rest.frame.relation}
\\
\nonumber
&=&
	\Bs
	\begin{array}[t]{l}	\displaystyle
		\int {d^4 p\over (2\pi)^4} {d^4 p'\over (2\pi)^4}
		\int {d^4 l\over (2\pi)^4} {d^4 l'\over (2\pi)^4}
	\\	\displaystyle
	\Rule{1.2\Tatescale}
	\times
		\tildebar{\Psi}[N^a_\alpha(P)](p,p')\;
		O^b_\mu(P,L;p,p',l,l')\;
		\Psi[N^c_\beta(L)](l,l')
	\end{array}
\\
\nonumber
&=&
	\Bs
	\begin{array}[t]{l}	\displaystyle
		\int {d^4 p\over (2\pi)^4} {d^4 p'\over (2\pi)^4}
		\int {d^4 l\over (2\pi)^4} {d^4 l'\over (2\pi)^4}
	\\	\displaystyle
	\Rule{1.2\Tatescale}
	\times
		\tildebar{\Psi}[N^a_\alpha(P)](p,p')\;
		O^b_\mu(P,L;p,p',l,l')\;
		\hat S(\Lambda^{-1})^{\otimes 3}
		\Psi[N^c_\beta(P)](\Lambda l,\Lambda l')
	\end{array}
\\
\nonumber
&=&
	\Bs
	\begin{array}[t]{l}	\displaystyle
		\int {d^4 p\over (2\pi)^4} {d^4 p'\over (2\pi)^4}
		\int {d^4 l\over (2\pi)^4} {d^4 l'\over (2\pi)^4}
	\\	\displaystyle
	\Rule{1.2\Tatescale}
	\times
		\tildebar{\Psi}[N^a_\alpha(P)](p,p')\;
		O^b_\mu(P,L;
			p,p',\Lambda^{-1} l,\Lambda^{-1}l')\;
		\hat S(\Lambda^{-1})^{\otimes 3}
		\Psi[N^c_\beta(P)](l,l'),
	\end{array}
\end{eqnarray}
where    the  second  equality follows   from    the following   Lorentz
transformation property of the \threeqBS{} amplitude of the rest frame:
\begin{eqnarray}
\lefteqn{
	\left\langle 0 \left| \Tate
		T \psi(x_1)\psi(x_2)\psi(x_3)
	\right| N^c_\beta(\Lambda^{-1}P) \right\rangle
}
\label{boost.3q.bs.amplitude}
\\
\nonumber
&=&
	\left\langle 0 \left| \Tate
		\left[T \psi(x_1)\psi(x_2)\psi(x_3)\right]
		\check{S}(\Lambda^{-1})
	\right| N^c_\beta(P) \right\rangle
\\
\nonumber
&=&
	\hat S(\Lambda^{-1})^{\otimes 3}
	\left\langle 0 \left| \Tate
		T \psi(\Lambda x_1)\psi(\Lambda x_2)\psi(\Lambda x_3)
	\right| N^c_\beta(P) \right\rangle,
\end{eqnarray}
where  ${(\Lambda^{-1})_{\mu}}^{\nu} = {(e^{-\lambda})_{\mu}}^{\nu}$  is
the pure boost  matrix, $\check{S}(\Lambda^{-1})$ is the boost  operator
in  the   Fock    space,  $\hat  S(\Lambda^{-1})   =   e^{S(-\lambda)}$,
($\displaystyle              S(-\lambda)       =                {-i\over
4}\sigma_{\mu\nu}(-\lambda)^{\mu\nu}$)  the  boost  matrix  in the Dirac
bispinor     space,   and   $\hat    S(\Lambda)^{\otimes   3}    =  \hat
S(\Lambda)\otimes \hat S(\Lambda)\otimes  \hat S(\Lambda)$.  In the last
line of \Eq{rest.frame.relation}, with the aid of the Lorentz invariance
of the Jacobian, we rotate the integration variables $l,l'$ by the boost
$\Lambda$, i.e., $l \mapsto  \Lambda^{-1}l, l' \mapsto  \Lambda^{-1}l'$.
The problem thus   reduces  to the  ``matrix  element''   calculation of
$O^b_\mu(P,\Lambda^{-1}P;    p,p',  \Lambda^{-1}l,   \Lambda^{-1}l')\hat
S(\Lambda^{-1})^{\otimes 3}$ between two  \threeqBS{} amplitudes  in the
rest frame.

By using the  homogeneous Faddeev  equation \Eq{homogeneous.faddeev.eq},
the   Faddeev  amplitude is  expressed   by the \threeqBS{} amplitude as
follows:
\begin{equation}
    \psi[N^a_{\alpha}(\vec P)]
=
    K_3
    \Psi[N^a_{\alpha}(\vec P)].
\end{equation}
Hence, the Lorentz transformation  property of the \threeqBS{} amplitude
implies   the following Lorentz transformation  property  of the Faddeev
amplitude:
\begin{equation}
    \psi[N^a_{\alpha}(\Lambda^{-1} P)](x_1,x_2,x_3)
=
    \hat S(\Lambda^{-1})^{\otimes 3}
    \psi[N^a_{\alpha}(P)](\Lambda x_1,\Lambda x_2,\Lambda x_3).
\end{equation}
By repeating  almost the same arguments, we  are left with the following
expression of the  matrix element in terms  of the Faddeev amplitudes in
the rest frame:
\begin{eqnarray}
\lefteqn{
    \left\langle N^a_\alpha(P) \left|\Tate
	A^b_{\mu}(x=0)
    \right| N^c_\beta(L) \right\rangle
}
\label{matrix.element.rest.frame.faddeev}
\\\nonumber
&=&
    \Bs
    \begin{array}[t]{l}	\displaystyle
	\int {d^4 p\over (2\pi)^4} {d^4 p'\over (2\pi)^4}
	\int {d^4 l\over (2\pi)^4} {d^4 l'\over (2\pi)^4}
    \\\displaystyle \Rule{1.5\Tatescale}
	\times
	\bar\psi[N^a_\alpha(P)](p,p')
	O_{F;\mu}^b(P,L;p,p',\Lambda^{-1},\Lambda^{-1}l')
	\hat S(\Lambda^{-1})^{\otimes 3}
	\psi[N^c_\beta(P)](l,l'),
    \end{array}
\end{eqnarray}
where    the    Fourier    transforms        $\psi[N^c_\beta(P)](p,p')$,
$\bar\psi[N^a_\alpha(P)]  (p,p')$  are defined from  $\psi[N^c_\beta(P)]
(x_1,x_2,x_3)$,  $\bar\psi[N^a_\alpha(P)]   (x_1,x_2,x_3)$   similar  to
\Eq{fourier.transform}, and   $O_{F;\mu}^b  (P,L;p,p',l,l')$  is defined
from:
\begin{equation}
    O_{F;\mu}^b(x_1,x_2,x_3;y_1,y_2,y_3)
\equiv
    \left[\Tate
    {
	i\delta K_F^{[e]}(x_1,x_2,x_3;y_1,y_2,y_3)
    \over
	\delta   a_{\mu}^b(x=0)
    }
    \right]^{[0]}.
\end{equation}

\section{Non-vanishing $g_{\rm ax}$}
\label{appendix.non-vanishing.gax.gtv.index}

\subsection{General Formalism}

The  aim of this   appendix is to  consider  how to  extract $g_{A}$ and
$\tilde g_{\pi  NN}$  in  the case  $g_{\rm  ax}  \neq 0$ providing   an
analytical  expression of the  GT violation $\Delta_G(m_0)$.  Instead of
repeating  the    argument  similar        to   the one      given    in
\Section{section.5.1}, which would become  quite lengthy, we derive  the
expression of the form  factors in a  different  manner, which would  be
easier for  the  readers to understand.

We first  have to  consider  functional  derivatives of  the constituent
quark propagator,  i.e., $\displaystyle \delta S_F/\delta a_{\mu}^b$ and
$\displaystyle  \delta  S_F/\delta  p^b$,   which are   obtained  as the
solutions       to       \Eq{functional.derivative.propagator}       and
\Eq{functional.derivative.propagator.pb}.      We  parameterize  $\delta
\Sigma/\delta a_{\mu}^b$ and $\delta \Sigma/\delta p^b$ as follows:
\begin{eqnarray}
\lefteqn{
    \int d^4z'
    e^{iq(z' - z)}
    \left[{\delta \Sigma^{[e]}(z') \over \delta a_{\mu}^b(z)}\right]
    ^{[0]}
}
\label{eq.125}
\\\nonumber
&=&
    H(q^2)
    \left( q^{\mu} \gamma_5 {\tau^b \over 2} \right)
+
    G_1(q^2)
    \left( {q^{\mu}\fslash{q} \over q^2} \gamma_5 {\tau^b \over 2} \right)
+
    G_2(q^2)
    \left( \gamma^{\mu} - {q^{\mu}\fslash{q}\over q^2} \right)
    \gamma_5 {\tau^b \over 2}
\\\nonumber
\lefteqn{
    \int d^4 z'
    e^{iq(z' - z)}    
    \left[{\delta \Sigma^{[e]}(z') \over \delta p^b(z)}\right]
    ^{[0]}
}
\\\nonumber
&=&
    I_1(q^2) \left(i \gamma_5 {\tau_b\over 2}\right)
+
    I_2(q^2) \left( i\fslash{q}\gamma_5 {\tau_b\over 2}\right).
\end{eqnarray}
$H(q^2)$, $G_1(q^2)$, $G_2(q^2)$,  $I_1(q^2)$ and $I_2(q^2)$ satisfy the
following coupled equations:
\begin{eqnarray}
    G_2(q^2)
&=&
    2ig_{\rm ax} \Pi_2(q^2)
+
    2ig_{\rm ax} \Pi_2(q^2)
    G_2(q^2)
\\\nonumber
    \left(
    \begin{array}{l}
	H(q^2) \\ G_1(q^2)
    \end{array}
    \right)
&=&
    \left(
    \begin{array}{l}
	-2ig_{\pi} \Pi_{5A}(q^2) \\ 2ig_{\rm ax} \Pi_{1}(q^2)
    \end{array}
    \right)
+
    \left(
    \begin{array}{ll}
	-2ig_{\pi} \Pi_{55}(q^2)
    &
	-2ig_{\pi} \Pi_{5A}(q^2)
    \\
	q^2 2ig_{\rm ax} \Pi_{5A}(q^2)
    &
	2ig_{\rm ax} \Pi_{1}(q^2)
    \end{array}
    \right)
    \left(
    \begin{array}{l}
	H(q^2) \\ G_1(q^2)
    \end{array}
    \right)
\\\nonumber
    \left(
    \begin{array}{l}
	I_1(q^2) \\ I_2(q^2)
    \end{array}
    \right)
&=&
    \left(
    \begin{array}{l}
	-2ig_{\pi} \Pi_{55}(q^2) \\ 2ig_{\rm ax} \Pi_{5A}(q^2)
    \end{array}
    \right)
+
    \left(
    \begin{array}{ll}
	-2ig_{\pi} \Pi_{55}(q^2)
    &
	q^2 (-2ig_{\pi}) \Pi_{5A}(q^2)
    \\
	2ig_{\rm ax} \Pi_{5A}(q^2)
    &
	2ig_{\rm ax} \Pi_{1}(q^2)
    \end{array}
    \right)
    \left(
    \begin{array}{l}
	I_1(q^2) \\ I_2(q^2)
    \end{array}
    \right),
\end{eqnarray}
with the following solutions:
\begin{eqnarray}
    H(q^2)
&=&
    {-2ig_{\pi} \Pi_{5A}(q^2) \over D(q^2)}
\label{eq.127}
\\\nonumber
    G_1(q^2)
&=&
    {
	q^2
	\left(\Tate 2ig_{\rm ax} \Pi_{5A}(q^2)\right)
	\left(\Tate -2ig_{\pi} \Pi_{5A}(q^2) \right)
    +
	\left(\Tate 1 + 2ig_{\pi} \Pi_{55}(q^2) \right)
	\left(\Tate 2ig_{\rm ax}\Pi_1(q^2) \right)
    \over
	D(q^2)
    }
\\\nonumber
    G_2(q^2)
&=&
    {2ig_{\rm ax} \Pi_2(q^2) \over 1 - 2ig_{\rm ax} \Pi_2(q^2)}
\\\nonumber
    I_1(q^2)
&=&
    {
	\left(\Tate 1 - 2ig_{\rm ax} \Pi_{1}(q^2) \right)
	\left(\Tate -2ig_{\pi} \Pi_{55}(q^2) \right)
    +
	q^2
	\left(\Tate - 2ig_{\pi} \Pi_{5A}(q^2) \right)
	\left(\Tate 2ig_{\rm ax} \Pi_{5A}(q^2) \right)
    \over
	D(q^2)
    }
\\\nonumber
    I_2(q^2)
&=&
    { 2ig_{\rm ax} \Pi_{5A}(q^2) \over D(q^2) }
\\\nonumber
    D(q^2)
&=&
    \left(\Tate 1 + 2ig_{\pi} \Pi_{55}(q^2)\right)
    \left(\Tate 1 - 2ig_{\rm ax} \Pi_{1}(q^2) \right)
-
    q^2
    \left(\Tate -2ig_{\pi} \Pi_{5A}(q^2) \right)
    \left(\Tate 2ig_{\rm ax} \Pi_{5A}(q^2) \right).
\end{eqnarray}
We  define  ``baryonic  parts'' $B_P(q^2)$ and   $B_A(q^2)$  of the form
factors through \Fig{fig.baryonic.pats}.
Now the matrix elements are expressed as follows:
\begin{eqnarray}
\lefteqn{
    \left\langle N(P) \left|\Tate
	A_{\mu}^b(x=0)
    \right| N(L) \right\rangle
=
    \Bar u(P)
    {\tau^b \over 2}
    \left(\Tate
	g_A(q^2) (\gamma_{\mu}\gamma_5)
    +
	h_A(q^2) (q_{\mu}\gamma_5)
    \right)
    u(L)
}
\\\nonumber
&=&
    \Bs
    \begin{array}[t]{l}	\displaystyle
	\left(\Tate
	    \Bar u(P) {\tau^b \over 2} \gamma_{\mu}\gamma_5 u(L)
	\right)
	B_A(q^2)
	\left(\Tate 1 + G_2(q^2) \right)
    \\\displaystyle\Tatetate
    +
	\left(\Tate
	    \Bar u(P) {\tau^b \over 2} \fslash{q} \gamma_5 u(L)
	\right)
	B_A(q^2)
	q_{\mu}
	{ G_1(q^2) - G_2(q^2) \over q^2 }
    \\\displaystyle\Tatetate
    +
	\left(\Tate
	    \Bar u(P) {\tau^b \over 2} \gamma_5 u(L)
	\right)
	B_P(q^2)
	q_{\mu} H(q^2)
    \end{array}
\\
\lefteqn{
    \left\langle N(P) \left|\Tate
	i \Bar\psi(0) \gamma_5 {\tau^b \over 2} \psi(0)
    \right| N(L) \right\rangle
=
    \Bar u(P)
    {\tau^b \over 2}
    \left(\Tate
	i_A(q^2) (i\gamma_5)
    \right)
    u(L)
}
\\\nonumber
&=&
    \left(\Tate
	\Bar u(P) {\tau^b \over 2} i\gamma_5 u(L)
    \right)
    B_P(q^2)
    \left(\Tate
	1 + I_1(q^2)
    \right)
+
    \left(\Tate
	\Bar u(P) {\tau^b \over 2} i\fslash{q} \gamma_5 u(L)
    \right)
    B_A(q^2)
    I_2(q^2),
\end{eqnarray}
where the isospin and  the helicity indices of the  initial and the final
nucleons are suppressed for simplicity.
The form factors are expressed as
\begin{eqnarray}
    g_A(q^2)
&=&
    B_A(q^2)\left(\Tate 1 + G_2(q^2)\right)
\label{three.form.factors}
\\\nonumber
    h_A(q^2)
&=&
    B_P(q^2) H(q^2)
+
    2m_N B_A(q^2)
    {G_1(q^2) - G_2(q^2) \over q^2}
\\\nonumber
    i_A(q^2)
&=&
    B_P(q^2) \left(\Tate 1 + I_1(q^2) \right)
+
    2m_N B_A(q^2) I_2(q^2).
\end{eqnarray}
$\tilde g_{\pi NN}$  is extracted according  to \Eq{extract.g.pi.nn} and
$g_A$ by $g_A = g_A(q^2 =  0)$.  Note that, due to \Eq{ward.identity.3},
$G_1(q^2) -  G_2(q^2)$ is proportional to  $q^2$,  which cancel $q^2$ in
the denominator in the expression of $h_A(q^2)$.

By   neglecting the    small cut-off    artifact,   the PCAC    relation
\Eq{desired.relation} leads to the relation:
\begin{equation}
    2m_N g_A(q^2) + q^2 h_A(q^2) = 2m_0 i_A(q^2),
\label{form.factor.satisfies}
\end{equation}
which provides the following relation between $B_P(q^2)$ and $B_A(q^2)$:
\begin{equation}
    2m_N B_A(q^2)
    \left(\Tate
	1 + G_1(q^2) - 2m_0 I_2(q^2)
    \right)
+
    B_P(q^2)
    \left(\Tate
	q^2 H(q^2)
    -	2m_0(1 + I_1(q^2))
    \right)
=
    0.
\label{relation.between.bp.and.ba}
\end{equation}
By            using   \Eq{eq.127},      \Eq{three.form.factors}      and
\Eq{relation.between.bp.and.ba},   we  are   left  with   the  following
analytical expression of $\Delta_{\rm G}(m_0)$:
\begin{eqnarray}
    \Delta_{\rm G}(m_0)
&=&
    {f_{\pi} \tilde g_{\pi NN} \over m_N g_A}
\label{gtv.mesonic.expression}
\\\nonumber
&=&
    {m_{\pi}^2 \over 2m_0}
    {-2ig_{\pi} \Pi_{5A}(0) \over 1 - 2ig_{\rm ax}\Pi_1(0)}
+
    m_{\pi}^2
    { 2ig_{\rm ax} \over 1 - 2ig_A \Pi_1(0)}
    \left(\Tate \Pi'_1(0) - \Pi'_2(0) \right),
\end{eqnarray}
where $\displaystyle \Pi'_1(q^2)   \equiv {d \Pi_1(q^2)\over   d(q^2)}$,
$\displaystyle \Pi'_2(q^2) \equiv {d  \Pi_2(q^2)\over d(q^2)}$.  All the
baryonic quantities disappear  from $\Delta(m_0)$.  Note  that, once the
mean-field approximation for the vacuum  is adopted, this expression  is
valid even beyond  the ladder truncation  scheme of the  3qBS kernel, as
far      as the    kernel      satisfies     our  criterion   of    PCAC
relation\footnote{Although    the    baryonic  parts   $B_{P}(q^2)$  and
$B_{A}(q^2)$ change, they still satisfy \Eq{relation.between.bp.and.ba}.
Hence  \Eq{gtv.mesonic.expression}  remains to  be   valid.}.   This, in
particular, implies that, to  improve $\Delta_{\rm G}(m_0)$, we  have to
improve the vacuum or to evaluate the on-shell $g_{\pi NN}$.  Therefore,
the easiest way to improve the result of $\Delta_{\rm G}(m_0)$ in
\Section{section.off.shell.gpinn.off.the.chiral.limit} is to improve the
vacuum by including the effect  of $g_{\rm ax} \neq 0$,  since it is not
so easy to treat the vacuum beyond the meanfield approximation.

\subsection{The Numerical Results}

We first have  to explain the choice  of the parameters.  There are five
parameters: the cutoff  $\Lambda$ (Euclidean sharp cutoff), the  current
quark mass $m_0$, the effective coupling constant  in the pionic channel
$g_{\pi}$,  the     effective coupling    constant   in   the iso-vector
axial-vector mesonic channel $g_{\rm  ax}$  and the  effective  coupling
constant  in  the $qq$ scalar  diquark channel  $g_{\rm sd}$.   We treat
$g_{\rm  ax}$ as a  free parameter, and fix   the first three parameters
($\Lambda,     m_0,     g_{\pi}$)   by   solving   the      gap equation
\Eq{explicit.gap.equation}  for   the constituent  quark  mass  $M$, and
\Eq{vacuum.parameter.2} for  the pion mass  $m_{\pi}$ and the pion decay
constant  $f_{\pi}$   imposing the     following three   conditions: (1)
$m_{\pi}= 140$ MeV, (2) $f_{\pi} =  93$ MeV, (3) $M  = 400$ MeV.  We fix
$g_{\rm sd}$ by requiring that $m_{N} = 940$ MeV.  We give our numerical
results  for the three  cases (1) $g_{\rm  ax}/g_{\pi} = 0$, (2) $g_{\rm
ax}/g_{\pi}  =  0.25$, (3) $g_{\rm  ax}/g_{\pi}   = 0.5$.  The  explicit
values of the parameters are listed as follows:
\begin{enumerate}
\item $g_{\rm ax}/g_{\pi} =  0$: $\Lambda = 739$  MeV, $m_0 = 8.99$ MeV,
$g_{\pi} = 10.4$ GeV$^{-2}$, $g_{\rm s}/g_{\pi} = 0.66$.
\item $g_{\rm ax}/g_{\pi}  =  0.25$: $\Lambda =  812$ MeV,  $m_0 = 6.99$
MeV, $g_{\pi} = 8.13$ GeV$^{-2}$, $g_{\rm s}/g_{\pi} = 0.694$.
\item $g_{\rm ax}/g_{\pi} = 0.5$: $\Lambda = 874$ MeV, $m_0 = 5.75$ MeV,
$g_{\pi} = 6.71$ GeV$^{-2}$, $g_{\rm s}/g_{\pi} = 0.72$.
\end{enumerate}
Once these parameters are fixed, we  consider the chiral limit by taking
$m_0 \to 0$ keeping $\Lambda$, $g_{\pi}$,  $g_{\rm ax}$ and $g_{\rm sd}$
fixed.  \Eq{explicit.gap.equation} provides  the $m_0$ dependences of $M
=  M(m_0)$.  We confine ourselves   to  non-negative values of   $g_{\rm
ax}/g_{\pi}$  because of the following  reasons: (1) our examples of the
interaction   Lagrangians   give  non-negative  values,   i.e.,  $g_{\rm
ax}/g_{\pi} = 0$  in the original   NJL type, and  $g_{\rm ax}/g_{\pi} =
0.5$ in the color current type.  (2) As we shall  see below, $g_{A}$ and
$\tilde  g_{\pi NN}$ increase with decreasing  $g_{\rm ax}/g_{\pi}$.  If
$g_{\rm  ax} < 0$,  we cannot  adjust  either $g_{A}$ nor $\tilde g_{\pi
NN}$.  (3) The  increase of $g_{A}$  with decreasing $g_{\rm ax}$ is due
to the  fact   that  the  negative   $g_{\rm  ax}/g_{\pi}$  works  as an
attraction in the  transversal  iso-vector axial-vector mesonic  channel
($a_1$ channel),  which makes $G_2(q^2=0)$   to grow up, leading  to the
rapid increase of  $g_A=g_A(q^2 =  0)$.  (cf.   \Eq{three.form.factors})
This   ``anti-screening''   of   $g_A$     does   not   seem    to    be
reasonable\cite{vogl.weise}.

Now  we present our numerical  results.  In \Fig{FigA1}, we plot $g_{\pi
NN}$  against  the current  quark mass $m_0$  for   these three cases of
$g_{\rm  ax}$.  It  is   seen that $\tilde  g_{\pi  NN}$  decreases with
increasing $g_{\rm  ax}/g_{\pi}$,  and increases  with increasing $m_0$.
The crosses in the figure are used to indicate $m_0$ which correspond to
$m_{\pi}  = 140$  MeV.  We  use  the diamonds   to indicate $m_0$  which
correspond to $m_{\pi} = 2\times 140$ MeV, which is used to indicate the
validity of the  single  pole dominance approximation of  $\tilde g_{\pi
NN}$.   Note that the distance between  $0$ and $4M^2$, i.e., the $q\bar
q$ cut, is still $10$ times larger than $m_{\pi}^2$, as long as $m_{\pi}
\le 2\times 140$ MeV.
In \Fig{FigA2}, we  plot  $g_{A}$ against  $m_0$ for the  three cases of
$g_{\rm  ax}$.  We see   that  $g_A$ decreases with increasing   $g_{\rm
ax}/g_{\pi}$ and increases with increasing $m_0$.
Note  that,  if $g_{\rm  ax}/g_{\pi} =   0.1$, then $g_{A}  =  1.23$ and
$\tilde g_{\pi NN} = 12.5$.  So the best fit of  $\tilde g_{\pi NN}$ and
$g_A$ could be obtained in the region $0 \le  g_{\rm ax}/g_{\pi} < 0.1$,
which, however, would depend on the quantity which  we prefer to adjust.
From this  point   of view, $g_{\rm   ax}=0$ is  actually a rather  good
choice.  In this case, $\tilde g_{\pi NN} = 13.5$ is very reasonable and
$g_{A}  = 1.33$ is still   close  to the experimental value  $g_{A}^{\rm
(exp)} = 1.26$.  Note that, due to the chiral symmetry, positive $g_{\rm
ax}/g_{\pi}$ implies an  attractive interaction in the iso-vector vector
mesonic channel ($\rho$ meson). In order to describe $\rho$ meson in the
NJL     model,    however,   we        need   stronger   $g_{\rm    ax}$
\cite{vogl.weise,takizawa}, but, stronger  $g_{\rm ax}$ leads to smaller
$g_A$ and $\tilde g_{\pi NN}$  in our calculation.  (The situation could
be improved  by including $qq$  interactions in the axial-vector diquark
channel, which enhance $g_{\pi  NN}$ \cite{alkofer}.  In this  case, one
should also include  the $qq$ interaction in  the vector diquark channel
for the Goldberger-Treiman and the PCAC relation.)
In \Fig{FigA3}, we plot the numerical $\Delta_{\rm G}$ together with the
analytic  $\Delta_{\rm G}$  (\Eq{gtv.mesonic.expression}) against $m_0$.
It is  seen that  they  are monotonically decreasing  functions of $m_0$
independent of $g_{\rm    ax}/g_{\pi}$.  We  see  that the   discrepancy
between the analytic  and the numeric  $\Delta_{\rm G}$ is within 3  \%.
Recall that, when deriving \Eq{gtv.mesonic.expression}, we neglected the
small cut-off  artifact.  Hence this  discrepancy  is solely due to  the
cut-off  artifact.  Note that   the analytic $\Delta_{\rm G}$ approaches
$1$  as    $m_0 \to     0$   in  \Fig{FigA3}.    In   fact,     based on
\Eq{ward.identity.1}, \Eq{ward.identity.2}, \Eq{ward.identity.3} and the
Gell-Mann-Oakes-Renner   relation,  it is possible  to    prove that the
analytic $\Delta_{\rm G}$ approaches to $1$ as $m_0 \to 0$.
To make   $\Delta_{\rm   G}$ an   increasing function   of $m_0$, it  is
necessary to  go beyond the   validity of the analytic  $\Delta_{\rm G}$
(\Eq{gtv.mesonic.expression}).  This implies   that all  we can do    is
either to improve  the gap equation for the  vacuum beyond the meanfield
approximation or to evaluate the on-shell $g_{\pi NN}$.
In  \Fig{FigA4}, we  plot the  PCAC  violation $\Delta_{\rm  P}$ against
$m_0$ for the  three cases of   $g_{\rm ax}/g_{\pi}$.  The  deviation of
$\Delta_P$ from $1$ is  solely due to the cutoff  artifact.  We see that
the  PCAC violations are only within   3\%, which suggests the practical
validity of our results.



\section*{Figure Captions}
\begin{enumerate}
\item \label{fig1} The  gauge transformation property  of $S_F^{[e]}$ is
depicted.   A single  line  represents a  propagator  of the constituent
quark in  the presence of the external  fields. The gauge transformation
reduces to the multiplication    of the phase factors   $\Omega(x)$  and
$\Omega(y)$.
\item \label{fig2}  The functional derivative   of the propagator, i.e.,
$\displaystyle \left[\Tate  {i\delta S_F^{[e]}\over \delta a_{\mu}^b(z)}
\right]^{[0]}$ is depicted.  A   single   line is a  constituent   quark
propagator. Dashed lines are used to indicate the momentum transfer $q$.
\Fig{fig2}.(a) corresponds to $g_{\rm  ax} = 0$ case, and \Fig{fig2}.(b)
corresponds to $g_{\rm ax} \neq 0$ case.
\item  \label{fig3} (a) The \threeqBS{} kernel  in the ladder truncation
scheme is  depicted.  A  single  line represents a  quark  propagator. A
slash indicates an amputation, i.e., removement of the quark leg.\\
(b) The diagram  used   to evaluate the    nucleon matrix  element   for
\threeqBS{}   kernel   of   the  type     \Fig{fig3}.(a)  is   depicted.
``$\otimes$'' is understood in the sense of \Fig{fig2}.  Diagrams of the
same topologies are omitted.
\item \label{fig4} (a)   The  $q\bar{q}$ exchange  improved  \threeqBS{}
kernel is depicted.   A single line  represents  a quark propagator.   A
slash indicates an  amputation.  A wavy line  represents a ladder sum of
$q\bar{q}$ bubbles  as is depicted in the  second line.  Diagrams of the
same topologies are omitted.\\
(b) The diagrams  with which to  obtain  the nucleon  matrix element for
\threeqBS{} kernel of the type \Fig{fig4}.(a)  are depicted.  The second
term     is often referred  to    as    the ``meson exchange   current''
contribution.  The diagrams of the same topologies are omitted.
\item  \label{fig5} The  proof   that the ladder truncated   \threeqBS{}
kernel gives  rise  to the PCAC  relation  correctly  is diagrammatically
explained.  In the first step, the local gauge transformation $\Omega(x)
= e^{i\gamma_5 \omega(x)}$  is applied.    Only the constituent    quark
propagators transform.  In the  next step, \Eq{chiral.invariance.diq} is
used.  The last step is  due to the fact  that amputated propagators are
delta functions. It is seen that the sufficient condition is satisfied.
\item  \label{fig6} The proof  that   the $q\bar{q}$ exchange   improved
\threeqBS{} kernel  gives  rise   to  the  PCAC  relation  correctly  is
diagrammatically explained.    In the first  step,  we applied the gauge
transformation.   In  the  second  step, \Eq{chiral.invariance} is used.
Note that all the phase factors, which  appear at the internal vertices,
cancel themselves.  The last step is due  to the fact that the amputated
propagators  are  delta  functions.  It  is   seen  that the  sufficient
condition is satisfied.
\item         \label{fig.faddeev}      (a)    The    Faddeev    equation
\Eq{faddeev.eq.for.g3} is depicted. A single line is a constituent quark
propagator.   A    double line is  a     t-matrix in the    $qq$ diquark
channel. (For  precise  meaning, see  \Fig{fig.faddeev}.(b).)   (b)  The
diagramatic  expression  of the  two-quark resolvent $\displaystyle {K_3
\over  1  - K_3}$ is  depicted.  A  single  line is  a constituent quark
propagator. A slash indicates an amputation.
\item  \label{fig7} The diagrams which contribute  to the matrix element
$\langle  N| A_{\mu}^b(x) |  N\rangle$ are depicted.   A blob with ``N''
which is followed by a triple  line is a quark-diquark Faddeev amplitude
of the nucleon.  A single line is  a constituent quark propagator, and a
double line is a t-matrix in the  $qq$ scalar diquark channel.  A dashed
line  indicates a   momentum  transfer   $q_{\mu}$.   \Fig{fig7}.(a)  is
referred to  as the  ``quark current'' contribution,  and \Fig{fig7}.(b)
the ``exchange current'' contribution.  \Fig{fig7}.(c) is referred to as
the  ``diquark current''  contribution.     If the $qq$  interaction  is
truncated   to the   scalar   diquark channel,  \Fig{fig7}.(c)  does not
contribute  to $g_{A}$,  due   to the iso-scalar  nature  of  the scalar
diquark.  \Fig{fig7}.(a')  is  equivalent  to  \Fig{fig7}.(a), which  is
obtained by once iterating the Faddeev  amplitude in the finial state by
using  the      homogeneous    Faddeev     equation.         By    using
\Eq{matrix.element.rest.frame.faddeev}       and   $\gamma^{\mu}(\Lambda
p)_{\mu} =   S(\Lambda)\fslash{p}S(\Lambda^{-1})$,   \Fig{fig7}.(a') and
\Fig{fig7}.(b) reduce to the four diagrams depicted in \Fig{fig7}.(d).
\item \label{fig8} Diagrams which  contribute to $g_A$ and  $g_{\pi NN}$
are  depicted.  These eight diagrams  are equivalent to four diagrams in
\Fig{fig7}.(d) in the limit $q  \to 0$, which  is indicated by ``$q  \to
0$''.  The explicit  spin-parity projection  shows  that the  first  two
contribute to $g_A$, and the others to $g_{\pi NN}$.
\item  \label{FigN1} The nucleon  mass  $m_N$ (for $g_{\rm sd}/g_{pi}  =
0.66$) (solid  line),  the  quark-diquark  threshold $M  +  m_{\rm sd}$
($g_{\rm sd}/g_{\pi}   =  0.66$  case)  (dotted line),  the   pion  mass
$m_{\pi}$ (dashed line), the pion  decay constant $f_{\pi}$  (dot-dashed
line), the constituent quark mass $M$  (dot-dot-dashed line) are plotted
against the current  quark mass $m_0$.  The  points which corresponds to
$m_{\pi} = 140$ MeV are indicated by the vertical dotted line.
\item \label{FigN2}  The nucleon     mass is plotted     against $g_{\rm
sd}/g_{\pi}$  for the two  cases (1) the  chiral limit (solid line), (2)
off the   chiral limit {}[{}$m_{\pi}  =   140$ MeV] (dashed   line). The
vertical   dotted line  indicates   $g_{\rm sd}/g_{\pi}  = 0.66$,  where
$m_{N}$ becomes $940$ MeV for the case $m_{\pi} = 140$ MeV.
\item \label{FigN3} $g_{\pi NN}$ is plotted against $g_{\rm sd}/g_{\pi}$
for the two cases (1) the chiral limit  (solid line), (2) off the chiral
limit {}[{}$m_{\pi} = 140$ MeV] (dashed line).  The vertical dotted line
indicates $g_{\rm  sd}/g_{\pi} = 0.66$, where  $m_{N}$ becomes $940$ MeV
for the case $m_{\pi} = 140$ MeV.
\item \label{FigN4} The iso-vector   $g_{A}$ is plotted  against $g_{\rm
sd}/g_{\pi}$ for the  two cases (1) the  chiral limit (solid line),  (2)
off   the chiral limit  {}[{}$m_{\pi} =  140$  MeV]  (dashed line).  The
vertical    dotted line indicates   $g_{\rm   sd}/g_{\pi} = 0.66$, where
$m_{N}$ becomes $940$ MeV for the case $m_{\pi} = 140$ MeV.
\item \label{FigN5} The violation of the GT relation $\Delta_{\rm G}$ is
plotted against $g_{\rm  sd}/g_{\pi}$ for the  two cases  (1) the chiral
limit (solid line),  (2) off the chiral limit  {}[{}$m_{\pi} = 140$ MeV]
(dotted line).  The dashed line is the plot of the violation of the PCAC
relation $\Delta_{\rm  P}$ off  the  chiral limit.  The  vertical dotted
line indicates $g_{\rm sd}/g_{\pi} = 0.66$,  where $m_{N}$ becomes $940$
MeV for the case $m_{\pi} = 140$ MeV.
\item  \label{FigN6} The violation of  the GT  relation $\Delta_{\rm G}$
(solid line) and the violation   of the PCAC relation $\Delta_{\rm   P}$
(dashed line) are plotted against  the current quark  mass $m_0$ for the
case $g_{\rm sd}/g_{\pi} =   0.66$.  The vertical dotted  line indicates
the current quark mass $m_0$ which corresponds to $m_{\pi} = 140$ MeV.
\item \label{fig.fierz}  The   interaction Lagrangian  ${\cal  L}_I$  is
depicted in \Fig{fig.fierz}.(a). The diagramatic interpretations of this
interaction in the $q\bar{q}$ channel are classified into the two types:
the  direct  channel [\Fig{fig.fierz}.(b)]  and  the   exchange  channel
(\Fig{fig.fierz}.(c)).   \Fig{fig.fierz}.(b)  is    obtained        from
\Fig{fig.fierz}.(a) by using the Fierz identity.  Another Fierz identity
leads to the diagramatic  representation of the  interaction in the $qq$
channel which is depicted in \Fig{fig.fierz}.(d).
\item   \label{fig.baryonic.pats}  The  baryonic  parts  $B_P(q^2)$  and
$B_A(q^2)$ of the form factors are defined diagrammatically.  A blob with
``N'' which is   followed by a  triple  line is a quark-diquark  Faddeev
amplitude  of  the nucleon.   A   single line is   the constituent quark
propagator, and a double line  is a t-matrix  in the $qq$ scalar diquark
channel.  A dashed line indicates a momentum transfer $q_{\mu}$.
\item  \label{FigA1} $g_{\pi NN}$ is plotted  against  the current quark
mass   $m_0$ for the  three cases  (1) $g_{\rm  ax}/g_{\pi}  = 0$ (solid
line),  (2) $g_{\rm  ax}/g_{\pi}  = 0.25$   (dotted line),  (3)  $g_{\rm
ax}/g_{\pi} = 0.5$ (dashed line).  The crosses indicate the points which
correspond to $m_{\pi} = 140$ MeV, and the  diamonds indicate the points
which correspond to $m_{\pi} = 280$ MeV.
\item  \label{FigA2} $g_{A}$ is plotted  against  the current quark mass
$m_0$ for the three cases (1) $g_{\rm ax}/g_{\pi} = 0$ (solid line), (2)
$g_{\rm ax}/g_{\pi}  =  0.25$ (dotted  line), (3) $g_{\rm  ax}/g_{\pi} =
0.5$ (dashed line).  The crosses indicate the points which correspond to
$m_{\pi} =  140$  MeV, and   the  diamonds  indicate the  points   which
correspond to $m_{\pi} = 280$ MeV.
\item  \label{FigA3} The numerical  $\Delta_{G}$  is plotted against the
current quark  mass $m_0$ for the three  cases (1) $g_{\rm ax}/g_{\pi} =
0$ (solid   line), (2) $g_{\rm ax}/g_{\pi} =   0.25$  (dashed line), (3)
$g_{\rm  ax}/g_{\pi}  =  0.5$    (dot-dot-dashed line).   The   analytic
$\Delta_{\rm G}$ is  also plotted for   these three cases (1) by  dotted
line, (2) by dot-dashed line,  (3) by dot-dash-dashed line.  The crosses
indicate  the points which correspond  to $m_{\pi}  =  140$ MeV, and the
diamonds indicate the points which correspond to $m_{\pi} = 280$ MeV.
\item \label{FigA4} The violation of the  PCAC relation $\Delta_{\rm P}$
is plotted against the current quark mass  $m_0$ for the three cases (1)
$g_{\rm  ax}/g_{\pi} = 0$ (solid  line), (2) $g_{\rm ax}/g_{\pi} = 0.25$
(dotted  line), (3)  $g_{\rm  ax}/g_{\pi}  = 0.5$   (dashed  line).  The
crosses indicate the points which correspond to $m_{\pi} = 140$ MeV, and
the diamonds indicate the points which correspond to $m_{\pi} = 280$ MeV.
\end{enumerate}

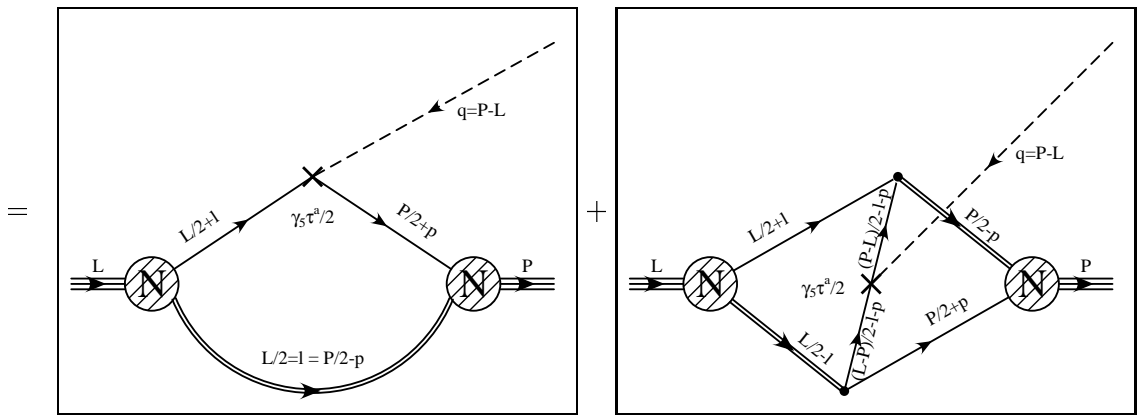
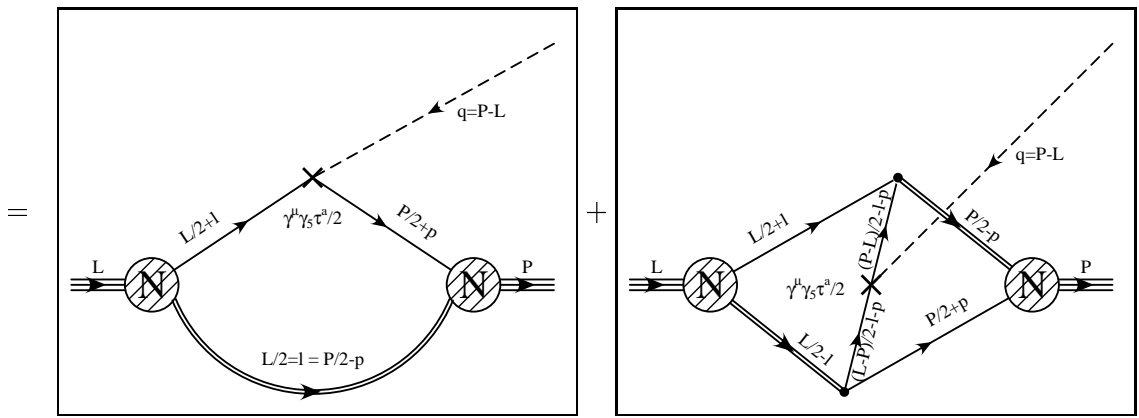
\end{document}